\documentclass[12pt,a4paper]{article}
\usepackage{jheppub_mod}
\usepackage[utf8]{inputenc}
\usepackage[T1]{fontenc}
\usepackage{amsmath,amsfonts,amssymb,physics,subcaption}
\usepackage{mathtools}
\usepackage[usenames,dvipsnames]{xcolor}
\usepackage{setspace}
\usepackage{verbatim}
\usepackage{xcolor, colortbl}
\usepackage{graphicx,cleveref}
\usepackage{empheq}
\usepackage{tikz}
\usepackage{tikz-cd} 
\usepackage{diagbox}
\usepackage{caption}    
\usepackage{varwidth}
\usetikzlibrary{arrows.meta,positioning,calc}

\newcommand{\circled}[1]{%
\tikz[baseline=(char.base)]{
\node[shape=circle,draw,inner sep=1pt] (char) {\small #1};}}

\def\mc{\mathcal}

\def\w[#1]{\widehat{#1}}
\def\vs[#1,#2]{\boldsymbol{{#1}_{#2}}}
\def\mes[#1]{d^{3}{#1}}
\def\del{\partial}

\def\<{\langle}
\def\>{\rangle}
\def\vecs[#1,#2]{\boldsymbol{{#1}_{#2}}}

\newcommand{\be}[1]{\begin{equation}\label{#1} }
\newcommand{\ee}{\end{equation}}
\newcommand{\bea}[1]{\begin{eqnarray}\label{#1} }
\newcommand{\eea}{\end{eqnarray}}

\renewcommand{\>}{\rangle}

\renewcommand{\a}{\alpha}

\newcommand{\s}{\sigma}

\usepackage{scalerel}

\def\a{\alpha}

\def\l{\lambda}

\def\m{\mu}
\def\n{\nu}
\def\N{\nabla}

\def\s{\sigma}

\usepackage{orcidlink}

\begin{document}
\title{A unified expansion of Einstein's gravity}
\author[\spadesuit]{Arkachur Bhattacharya\orcidlink{0009-0004-3153-6470}}
\author[\spadesuit]{,Pushkar Soni\orcidlink{0009-0001-8514-0859}}

\affiliation[\spadesuit]{Indian Institute of Technology Kanpur, Kanpur 208016, India.}

\emailAdd{arkachurb25@iitk.ac.in, pushkars21@iitk.ac.in }


\abstract{Non-Lorentzian theories of gravity, most common of which are Galilean and Carrollian gravity, arise from General relativity under suitable scalings. General relativity can be obtained by gauging the Poincar\'e algebra. A convenient formulation of non-Lorentzian gravity follows the contraction of the Poincar\'e algebra in the tangent space to its non-Lorentzian counterparts, e.g. Galilean and Carrollian algebras. In existing literature, different non-Lorentzian theories of gravity have been addressed separately. In this paper, we introduce a single unified framework of expansion to address all these different theories. We show that different scalings can be unified into a single covariant form parametrized by $(s,n)$, alongside the contraction parameter $\epsilon$. Keeping these parameters unfixed in the limit $\epsilon \to 0$ defines a $\textit{unified flat geometry}$ and its \textit{unified algebra}, which reduces to a specific non-Lorentzian geometry for a particular choice of $s$ and $n$. Using this setup in the tangent space, we systematically expand the Einstein-Hilbert action in even powers of $\epsilon$, which we call a \textit{unified expansion} of Einstein's gravity, whose leading-order theory is fixed by $(s,n)$. This reproduces various classes of gravitational theories, including Einstein's gravity (the trivial case), Galilean gravity, Carroll gravity, all of which can be extracted from this expansion. Using the expansion, we then formulate String Carroll (SC) gravity, where the local metric has two vanishing eigenvalues. The near-horizon region of generic non-extremal black holes has been recently shown to be an SC geometry. By considering explicit examples, we confirm that these near-horizon geometries constitute solutions of SC gravity, paving the way of understanding physics near the horizon of generic black holes in terms of SC gravity.}


\maketitle

\section{Introduction}
\label{sec:introduction}
With the advent of special and general relativity, there arose a sustained interest in exploring theories that are covariant under these relativity principles. More than a century later, this has remained a central theme in theoretical physics. Special relativity of course comes hand in hand with Lorentz symmetry and general relativity locally reduces to special relativity. So we will call them Lorentzian theories. More recently, attention has expanded beyond this Lorentzian paradigm and there is a growing interest in exploring theories that are invariant under different set of symmetries. They are collectively known as non-Lorentzian theories (we refer the reader to \cite{Bergshoeff:2022eog} for some idea of the recent development in this field). Among the most well-studied examples of non-Lorentzian theories are Galilean and Carrollian theories\footnote{We refer readers to the following reviews on Carroll physics \cite{Bagchi:2025vri,Ruzziconi:2026bix,deBoer:2023fnj}.}. Such theories have attracted considerable attention due to their appearance in diverse physical settings, including gravity and cosmology \cite{Hartong:2015zia,VandenBleeken:2017rij,Guerrieri:2021cdz,Hansen:2021fxi,March:2024zck,cardona2025higherordernewtoncartangravity,Concha:2025vhd,deBoer:2021jej,Afshar:2025imp}, field theory \cite{Bagchi:2009ca,Bagchi:2017yvj,Bagchi:2019xfx,Saha:2022gjw,Bagchi:2022eav,Cotler:2024xhb}, string theory and supergravity \cite{Gomis:2000bd,Bergshoeff:2018yvt,Gomis:2019zyu,Bagchi:2019cay,Bergshoeff:2023fcf,Bagchi:2026wcu,Bagchi:2026iyu,Bergshoeff:2021bmc,Lescano:2025ixp,Lescano:2026dhv,Ballesteros:2026bqe}, hydrodynamics \cite{Jensen:2014wha,Ciambelli:2018xat,Bagchi:2023ysc,Bagchi:2023rwd,Kolekar:2024cfg,Shukla:2026xig,Chabirand:2026sxe}, condensed-matter systems \cite{Bidussi:2021nmp,Bagchi:2022eui,Figueroa-OFarrill:2023vbj,Ara:2024fbr,Biswas:2025dte,Banerjee:2026kav}, and flat-space holography \cite{Bagchi:2010zz,Bagchi:2016bcd, Bagchi:2022emh, Donnay:2022aba, Bagchi:2023cen,Bagchi:2023fbj, Nguyen:2025zhg}, among others. In what follows, we concentrate on one particular setting among these --- a theory of gravity.

\medskip

Throughout the work, we will consider theories of gravity which arise from Einstein's theory of gravity{\footnote{We will, e.g., not consider higher curvature corrections.}}. A first-order formalism of Einstein's gravity relies on the fact that the spacetime manifold is locally flat (Minkowskian) and the local symmetry algebra is the Lorentz algebra\footnote{Although the actual local algebra is the Poincar\'e algebra, but local translations are realized as diffeomorphisms in a theory of gravity.}. Thus general relativity (GR) is regarded as a Lorentzian theory of gravity. We are interested in investigating theories of gravity in which the local symmetry algebra is modified by a contraction implemented on the local metric (for an example, refer to \cite{Duval:2014uoa}). Collectively, such theories are referred to as non-Lorentzian theories of gravity. Important examples include Galilean gravity \cite{a,b,c} and Carroll gravity \cite{Hartong:2015xda,Hansen:2021fxi,March:2024zck}. 

\medskip

A robust approach to investigate such theories is from the perspective of gauge theories, where the Poincar\'e group and its non-Lorentzian counterparts serve as the underlying gauge symmetry. This provides a systematic dictionary of all relevant fields and their transformations, along with their various interplays. One may then construct an invariant functional of these fields that serves as an action for a particular gravitational theory. Such examples of gravity as a gauge theory can be found in \cite{Hartong:2015xda,Hartong:2015zia,Obukhov:2018bmf,Obukhov:2022khx}\footnote{For a pedagogical review on gravity as a gauge theory, we refer readers to \cite{Bennett:2021dbg}.}. An alternative route is to directly expand the geometry and the Lorentzian theory of gravity in some ``smallness'' parameter $\epsilon$\footnote{A priori, this smallness parameter has not been assigned any direct physical meaning. It merely acts as a parameter attached to certain eigenvalues of the local metric such that, in the limit $\epsilon \to 0$, the corresponding scaling eigenvalues become suppressed relative to the non-scaling ones. }, which is responsible for modifying the local symmetry algebra, when $\epsilon\to0$ limit is imposed. In the present work, throughout we adopt the method of expansion and restrict ourselves to GR\footnote{In principle, this construction can also be extended to higher-derivative theories. This might become relevant once the low-energy approximation of quantum gravity ceases to be valid.}. One may then verify, through an intrinsic analysis, whether the resulting expansion consistently reproduces the desired non-Lorentzian geometric structure or not. Such expansions of general relativity leading to Galilean and Carroll gravity have already been done in \cite{c,Hansen:2021fxi,Elbistan:2022plu,Bal:2026xup}.

\medskip

But so far, in the literature, expansions have been carried out independently, each tailored to produce a specific theory of gravity at the leading order. In this paper, we demonstrate how a single unified expansion can be performed to generate various non-Lorentzian counterparts of the Einstein-Hilbert action, including both Galilean and Carroll gravity (see a similar expansion in \cite{Zorba:2026ghe}). This is made possible by a key feature of the local metric. In a suitable local frame, it admits a block-diagonal decomposition into scaling and non-scaling sectors, while leaving both the size of the scaling submatrix, denoted by `$n$', and the placement of the timelike signature unfixed. The latter may lie in either the scaling or the non-scaling sector, and we introduce the parameter `$s$' to distinguish between them (details on these parameters are discussed in \cref{sec:Symmetries}). We classify their choice in four distinct classes: 
\begin{itemize}
    \item [i.] Galilean class, 
    \item[ii.] Carrollian class, 
    \item[iii.] Lorentzian class and 
    \item[iv.] Euclidean class\footnote{In \cite{Bergshoeff:2023rkk,Bergshoeff:2026cxt}, geometries lying in the Galilean and Carrollian classes are called \textit{p}-brane Galilei and \textit{p}-brane Carroll geometry, respectively.}.
\end{itemize}
This enables us to carry out an expansion in a unified manner, keeping these choices arbitrary throughout the intermediate calculations and fixing them only at the end in order to recover a particular theory of gravity. 

\medskip

Rather than proceeding directly to the expansion, we first establish a unified framework for the local metric and its symmetry. In the limit $\epsilon\to0$, we call it a \textit{unified flat geometry} and a \textit{unified algebra}. We then generalize it to a generic background and call it a \textit{unified geometry}. Later, upon fixing the pair $(s,n)$, the discussion reduces to a specific local flat geometry and its associated isometry algebra. This choice simultaneously induces a corresponding interpretation of the expansion parameter $\epsilon$, geometry, and its dynamical theory.\\
\newline
Throughout, we adopt a similar systematics of expansion followed in \cite{c,Hansen:2021fxi}. The unified expansion is then performed as follows:
\begin{itemize}
    \item We begin by splitting the local metric (denoted by $\widetilde\eta$) into the scaling and the non-scaling sector while keeping the pair $(s,n)$ unfixed
    \begin{eqnarray}
\widetilde\eta=\epsilon^2\eta_{\text{scaling}}+ \eta_{\text{non-scaling}}.
    \end{eqnarray} 
  \item We then express the spacetime metric $g$ (inverse $g^{-1}$) in terms of the vielbeins, through which the scaling gets naturally induced
  \begin{eqnarray}
    g=\epsilon^2g_{\text{scaling}}+g_{\text{non-scaling}}.
  \end{eqnarray}
  \item We then postulate an expansion of the frame fields (vielbeins) in powers of $\epsilon$. This further triggers an expansion of metric, and its inverse in powers of $\epsilon$
  \begin{eqnarray}
      g=g_{(0)}+\epsilon^2 g_{(2)}+\mathcal{O}(\epsilon^4).
  \end{eqnarray}
  \item However, instead of implementing the expansion immediately, we keep it intact until the end. We first compute the relevant curvature scalars for the gravity theory and organize them order by order in overall $\epsilon$ factors, which we refer to as a \textit{unified decomposition}. The key feature of this procedure is that it allows the leading-order contribution to be extracted directly from the leading term in the decomposition. Subsequently, the higher-order contributions can also be extracted systematically.

  \item Finally, we perform the expansion. Then, by choosing different combinations of the pair $(s,n)$, the expansion specializes to different gravity theories at the leading order.
\end{itemize}
We can even fix this pair at the level of decomposition itself. As emphasized above, it directly allows one to extract the leading-order theory from the leading-order term in the decomposition.

\medskip 

In a nutshell, for Einstein's gravity we proceed as
\begin{figure}[h!]
    \centering
    \includegraphics[width=0.6\linewidth]{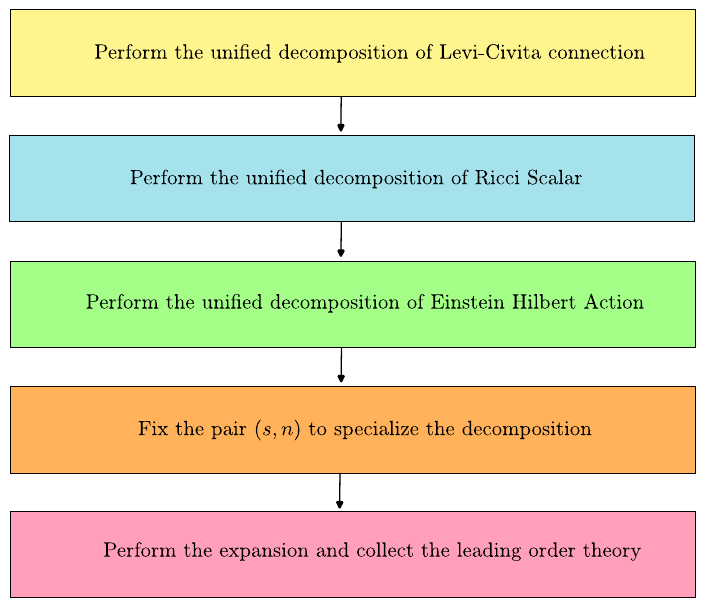}
    \label{fig:placeholder0}
\end{figure}

\medskip

This paper is arranged in the following manner. In \cref{sec:Symmetries}, we define the unified flat geometry and compute its unified symmetry algebra. This section contains the detailed role of the pair $(s,n)$. Then, in the next \cref{sec:Expansion-of-geometry}, we perform an expansion of Lorentzian geometry in powers of $\epsilon$ and show, in the limit $\epsilon\to0$ the unified geometry naturally emerges. In \cref {sec:Useful-connection}, we introduce a unified connection compatible with the unified geometry. After that, in \cref{sec:GR-expansion}, we perform a unified decomposition of the Einstein-Hilbert action. Subsequently, in \cref{sec:All-in one gravity}, we fix the pair $(s,n)$ and show how different theories of gravity can be systematically extracted from the unified expansion. In \cref{subsec: String Carroll gravity}, we separately discuss the case of string Carroll gravity along with some explicit examples. Finally, we conclude this paper with a discussion and an outlook towards problems that might be of interest for future work in \cref{sec:Discussion}. Appendices~\ref{app:derviation-of-connection}-\ref{appen:galilean gravity} provide additional material complementing the discussion in the main text.\\
\newline
\underline{\textit{Notation}:} Throughout, we work in a $d$-dimensional spacetime, unless mentioned otherwise. Greek indices $\mu,\nu,\ldots$ denote spacetime indices. While the local (tangent space) indices are split into two sets: capital Latin indices $A,B,\ldots$ label the scaling sector, while lowercase Latin indices $a,b,\ldots$ label the remaining non-scaling sector. 

Accordingly, in Lorentzian geometry, the local flat metric takes a block-diagonal form 
\begin{equation}
   \tilde \eta=\begin{pmatrix}
       \epsilon^2 \eta_{AB}&0_b\\
       0_a&\eta_{ab}
   \end{pmatrix},\quad \tilde \eta^{-1}=\begin{pmatrix}
       \epsilon^{-2} \eta^{AB}&0^b\\
       0^a&\eta^{ab}
   \end{pmatrix}.
\end{equation}
We use $\tilde\eta_{AB}$ and its inverse $\tilde \eta^{AB}$ to raise and lower capital Latin indices, while $\tilde \eta_{ab}$ and its inverse $\tilde \eta^{ab}$ are used for lowercase Latin indices. Whereas, the spacetime indices are raised and lowered using the full metric $g_{\mu\nu}$ and its inverse $g^{\mu\nu}$.

\section{Unified flat geometry and its symmetry}
\label{sec:Symmetries}
In this section, we systematize a unified framework that captures different classes of flat geometries (both Lorentzian and non-Lorentzian) and their associated symmetry algebras. We call it a \textit{unified flat geometry} and \textit{unified algebra}. It can be viewed as a local structure underlying a generic curved background. They are characterized by two parameters $s$ and $n$, such that different choices of these parameters pick out a particular flat geometry and its isometry algebra. We show that this unified structure admits four broader classifications: the Galilean, Carrollian, Lorentzian, and Euclidean classes.\\
\newline
$\circledast$\textit{For this section, we borrow the convention of local indices for spacetime indices}.

\medskip

\underline{\textit{Unified flat geometry and algebra}:} We start with the usual Minkowski line element on spacetime covered by Cartesian coordinates ($x^A,x^a$). However, we split this metric into the scaling and non-scaling sectors in a block diagonal form as
\begin{equation}\label{local-Lorentzian-flat}
   \tilde \eta=\begin{pmatrix}
       \epsilon^2 \eta_{AB}&0_b\\
       0_a&\eta_{ab}
   \end{pmatrix},\quad \tilde \eta^{-1}=\begin{pmatrix}
       \epsilon^{-2} \eta^{AB}&0^b\\
       0^a&\eta^{ab}
   \end{pmatrix}.
\end{equation}
where $\epsilon$ is the scaling parameter (which we later want to tune to zero), with  $\eta_{AB}$ and $\eta_{ab}$ as the scaling and non-scaling blocks, respectively. Before proceeding, let us first highlight a crucial point: in our convention, the ``Lorentzian'' flat geometry carries the following signature
\[
(-,\underset{d-1}{\underbrace{+,\cdots,+}}).
\]
The location of the negative eigenvalue, whether it lies in $\eta_{AB}$ or in $\eta_{ab}$, determines the interpretation of the resulting geometry and symmetry algebra when the $\epsilon\to0$ limit is imposed. We therefore consider the sub-metrics in the following form\footnote{It is important to note that this parametrization of the metric is a particular way to proceed. Alternative parametrizations may also be constructed, potentially leading to different yet equivalent notions of unified geometry and symmetry. }
\begin{eqnarray}
    \eta_{AB}=\begin{bmatrix}
        (-1)^s&0\\
        0&\mathbf{1}_{{(n-1)\times (n-1)}}
    \end{bmatrix}_{{n\times n}},\quad \eta_{ab}=\begin{bmatrix}
        (-1)^{1-s}&0\\
        0&\mathbf{1}_{{(d-n-1)\times (d-n-1)}}
    \end{bmatrix}_{{(d-n)\times (d-n)}}
\end{eqnarray}
where, $s$ takes value either 0 or 1. Here we have assumed that $\eta_{AB}$ is an $n$-dimensional matrix with $0\leq n\leq d$, and consequently $\eta_{ab}$ is a $(d-n)$-dimensional matrix. The pair ($s,n$) constitutes the two parameters of our construction. Different choices of these parameters select different geometries in the limit $\epsilon\to0$. Therefore we do not assume \textit{a priori} the location of the time-like signature and size of the sub-matrices, i.e. we won't fix the value of $s$ and $n$, respectively. Instead, we directly implement the contraction $\epsilon\to0$. Metric and its inverse contracts to
\begin{eqnarray}\label{unified-flat-geometry}
    h=\lim_{\epsilon\to0}\tilde \eta=\begin{pmatrix}
        0&0\\
        0&\eta_{ab}
    \end{pmatrix},\qquad v=\lim_{\epsilon\to0}\epsilon^2 \tilde\eta^{-1}=\begin{pmatrix}
        \eta^{AB}&0\\
        0&0
    \end{pmatrix}.
\end{eqnarray}
We call this data a \textit{unified flat geometry}\footnote{The case of $n=d$ needs to be handled with care. At first sight, the metric data appear to vanish in this limit. But it is crucial to note that the scaling parameter is set to zero only in the limiting sense and is never exactly zero. Its role is merely to control the relative magnitude of different metric components. In the case $n=d$, all components remain on the same footing, so the metric data in the strict sense does not vanish. Therefore, their finite values can be extracted by a rescaling. In any case, this subtlety does not affect our analysis. Since our ultimate goal is to perform an expansion in $\epsilon$'s, these factors only modify the overall scale of the expansion and do not alter its structure.}. It is defined via two sets of data, $h$ and $v$\footnote{We must note that $h$ and $v$ alone provide only a weaker definition of a unified flat geometry, because of the fiber bundle structure it induces on a manifold. Refer to Appendix \ref{app:fibre bundle} for details.}. Therefore, its isometry algebra is obtained by solving the following Killing conditions
\begin{eqnarray}
    \pounds_\xi h=0,\quad \pounds_\xi v=0.
\end{eqnarray}
The general solution reads
\begin{eqnarray}
    \xi^\mu\del_\mu=[\omega^{A}_{\ B}(x^a)x^B+c^A(x^a)]\del_A+[c^a+\omega^a_{\ b}x^b]\del_a,
\end{eqnarray}
where $\omega^{AB}$ and $\omega^{ab}$ are antisymmetric matrices\footnote{We have used the following notation to denote the raising and lowering of indices
\begin{eqnarray}
    \omega^{AB}=\omega^A_{\ C}\eta^{BC},\quad \omega^{ab}=\omega^{a}_{\ c}\eta^{bc}.
\end{eqnarray}}. The underdetermined functions $c^A(x^a)$ and $\omega^A_{\ B}(x^a)$ make the algebra infinite-dimensional. If we further demand that the algebra preserves the flat connection on the manifold (all components in an appropriate chart become zero), then the algebra reduces to its finite-dimensional sector, where $c^A$ is at most linear in $x^a$ and $\omega^{AB}$ is a constant. The corresponding generators thus become \footnote{These generators can directly be obtained from the Poincar\'e generators, by introducing the scale $x^A\to \epsilon x^A$ and then taking the limit $\epsilon\to0$.}
\begin{eqnarray}\label{unified-algbera}
     \widetilde P_A=\del_A,\quad \widetilde J_{aA}=x_a\del_A,\quad \widetilde J_{AB}=x_A\del_B-x_B\del_A,\quad \widetilde P_a=\del_a ,\quad \widetilde J_{ab}=x_a\del_b-x_b\del_a.\notag \\
\end{eqnarray}
We call this a \textit{unified algebra}. It satisfies the following non-trivial commutation relations
\begin{subequations}\label{n-commutations}
    \begin{align}
        &[\widetilde J_{ab},\widetilde{J}_{cA}]=2\eta_{c[b}\widetilde{J}_{a]A},~~[\widetilde J_{AB},\widetilde{J}_{aC}]=2\widetilde{J}_{a[A}\eta_{B]C},\\
        &[\widetilde J_{ab},\widetilde P_c]=2\eta_{c[b}\widetilde P_{a]},~~[\widetilde J_{AB},\widetilde{P}_C]=2\eta_{C[B}\widetilde{P}_{ A]},~~[\widetilde{J}_{aA},\widetilde P_b]=-\eta_{ab}\widetilde P_A,\\
        &[\widetilde J_{AB},\widetilde J_{CD}]=2\eta_{C[B}\widetilde J_{A]D}-2\eta_{D[B}\widetilde J_{A]C},\quad  [\widetilde J_{ab},\widetilde J_{cd}]=2\eta_{c[b}\widetilde J_{a]d}-2\eta_{d[b}\widetilde J_{a]c}.
    \end{align}
\end{subequations}
We now discuss the different classifications of the unified geometry and algebra by fixing the values for $s$ and $n$. While fixing the value of $s$, the scaling and non-scaling metric reduce to
\begin{subequations}
    \begin{align}
        &s=0:~~\eta_{AB}=\delta_{AB}, \quad \eta_{ab}=\bar\eta_{ab},\\
        &s=1:~~\eta_{AB}=\bar\eta_{AB},\quad \eta_{ab}=\delta_{ab},
    \end{align}
\end{subequations}
where $\bar\eta_{AB}$ and $\bar\eta_{ab}$ are Minkowskian metric. We find that the unified geometry and algebra naturally fall into four broad classes, as discussed below.\\
\newline
\textbf{Galilean class:} In this class, the time-like signature sits in $\eta_{ab}$, i.e. $s=0$. Hence, all indices with capital Latin indices are space-like, and one of the lowercase Latin indices is time-like. This enforces interpreting all generators in $\widetilde{J}_{AB}$ as spatial rotations and those in $\widetilde{J}_{ab}$ as Lorentz boosts plus rotations. Therefore, the unified Lie brackets in \cref{n-commutations} reduce to
\begin{subequations}\label{g-commutations}
    \begin{align}
        &[\widetilde J_{ab},\widetilde{J}_{cA}]=2\bar \eta_{c[b}\widetilde{J}_{a]A},~~[\widetilde J_{AB},\widetilde{J}_{aC}]=2\widetilde{J}_{a[A}\delta_{B]C},\\
        &[\widetilde J_{ab},\widetilde P_c]=2\bar\eta_{c[b}\widetilde P_{a]},~~[\widetilde J_{AB},\widetilde{P}_C]=2\delta_{C[B}\widetilde{P}_{ A]},~~[\widetilde{J}_{aA},\widetilde P_b]=-\bar\eta_{ab}\widetilde P_A,\\
        &[\widetilde J_{AB},\widetilde J_{CD}]=2\delta_{C[B}\widetilde J_{A]D}-2\delta_{D[B}\widetilde J_{A]C},\quad  [\widetilde J_{ab},\widetilde J_{cd}]=2\bar\eta_{c[b}\widetilde J_{a]d}-2\bar\eta_{d[b}\widetilde J_{a]c}.
    \end{align}
\end{subequations}
With this interpretation, we can identify $d-n$ copies of $(n+1)$-dimensional Galilean sub-algebras in \cref{unified-algbera}
\begin{equation}
    \text{Galilean}_a:\quad \{\widetilde P_a,~\widetilde P_A,~\widetilde J_{aA},~\widetilde J_{AB}\},
\end{equation}
where, $\widetilde P_a$ are Galilean time translations, $\widetilde P_A$ are spatial translations, $\widetilde J_{aA}$ are Galilean boosts and $\widetilde J_{AB}$ are spatial rotations. Each copy is labelled by the index $a$, while $\widetilde{P}_A$ and $\widetilde J_{AB}$ sit as common in each copy. It has two known special cases: i.  For $n=d-1$, full algebra reduces to a single copy of the $d$-dimensional Galilean algebra. In this case, `$a$' runs over a single index, the Galilean time index. Consequently, there are no remaining generators of the form $\widetilde{J}_{ab}$, and therefore, no Lorentz boost survives. ii. For $n=d-2$, we obtain two copies of a codimension one Galilean algebra and one Lorentz boost generator. In the literature, it is called string Galilei algebra \cite{Brugues:2004an,Andringa:2012uz,Bergshoeff:2019pij}. \\
\newline
\textbf{Carrollian class:} In this class, we place time-like signature in $\eta_{AB}$, i.e. $s=1$. Hence, all the lowercase Latin indices are now space-like, and one of the capital Latin indices is time-like. This makes generators in $\widetilde{J}_{AB}$ either a Lorentz boost or a spatial rotation, while $\widetilde{J}_{ab}$ only contains rotations. This choice reduces the commutations in \cref{n-commutations} to
\begin{subequations}\label{c-commutations}
    \begin{align}
        &[\widetilde J_{ab},\widetilde{J}_{cA}]=2\delta_{c[b}\widetilde{J}_{a]A},~~[\widetilde J_{AB},\widetilde{J}_{aC}]=2\widetilde{J}_{a[A}\bar\eta_{B]C},\\
        &[\widetilde J_{ab},\widetilde P_c]=2\delta_{c[b}\widetilde P_{a]},~~[\widetilde J_{AB},\widetilde{P}_C]=2\bar\eta_{C[B}\widetilde{P}_{ A]},~~[\widetilde{J}_{aA},\widetilde P_b]=-\delta_{ab}\widetilde P_A,\\
        &[\widetilde J_{AB},\widetilde J_{CD}]=2\bar\eta_{C[B}\widetilde J_{A]D}-2\bar\eta_{D[B}\widetilde J_{A]C},\quad  [\widetilde J_{ab},\widetilde J_{cd}]=2\delta_{c[b}\widetilde J_{a]d}-2\delta_{d[b}\widetilde J_{a]c}.
    \end{align}
\end{subequations}
Similar to the Galilean class, we can now identify $n$ copies of $(d-n+1)$-dimensional Carroll sub-algebras in \cref{unified-algbera}
\begin{equation}
    \text{Carroll}_A:\quad \{\widetilde P_a,~\widetilde P_A,~\widetilde J_{aA},~\widetilde J_{ab}\},
\end{equation}
where we now interpret $\widetilde P_A$ are Carroll time translations, $\widetilde P_a$ are spatial translations, $\widetilde J_{aA}$ as Carroll boosts and $\widetilde J_{ab}$ are spatial rotations. Here, each copy is labelled by an index $A$, while $\widetilde P_a$ and $\widetilde{J}_{ab}$ are common in each copy. This class contains two special cases: i. For $n=1$, we do not have any generators left in $\widetilde{J}_{AB}$. Hence, the full algebra reduces to the usual $d$-dimensional Carroll algebra, and ii. $n=2$ is the recently introduced string Carroll algebra \cite{Bagchi:2023cfp,Bagchi:2024rje}, which contains two codimension one copies of Carroll sub-algebras and one Lorentz boost generator (similar algebras were identified in \cite{Bagchi:2024epw,Majumdar:2024rxg}).\\
\newline
\textbf{Lorentzian class:} This class contains only two cases, namely $(s,n)=(1,d)$ and $(s,n)=(0,0)$. However, both cases are essentially trivial and equivalent, since they simply reproduce the $d$-dimensional Poincar\'e algebra, realized entirely within either the $\eta_{AB}$ sector or the $\eta_{ab}$ sector, respectively
\begin{subequations}
    \begin{align}
        &(s,n)=(1,d):~\{P_A,~J_{AB}\},\\
        &(s,n)=(0,0):~\{P_a,~J_{ab}\}.
    \end{align}
\end{subequations}
Therefore, we get either of the commutations
\begin{subequations}\label{l-commutations}
    \begin{align}
        &[\widetilde J_{AB},\widetilde{P}_C]=2\bar\eta_{C[B}\widetilde{P}_{ A]},\quad [\widetilde J_{AB},\widetilde J_{CD}]=2\bar\eta_{C[B}\widetilde J_{A]D}-2\bar\eta_{D[B}\widetilde J_{A]C},\\
        &\text{or},\quad [\widetilde J_{ab},\widetilde{P}_c]=2\bar\eta_{c[b}\widetilde{P}_{ a]},\quad [\widetilde J_{ab},\widetilde J_{cd}]=2\bar\eta_{c[b}\widetilde J_{a]d}-2\bar\eta_{d[b}\widetilde J_{a]c},
    \end{align}
\end{subequations}
respectively.\\
\newline
\textbf{Euclidean class:} Similar to the Lorentzian case, this class also contains only two possibilities, corresponding to  $(s,n)=(0,d)$ and $(s,n)=(1,0)$. This  again produces the full $d$-dimensional Euclidean algebra realized entirely in either the $\eta_{AB}$ or the $\eta_{ab}$ sector given by
\begin{subequations}
    \begin{align}
        &(s,n)=(0,d):~\{P_A,~J_{AB}\},\\
        &(s,n)=(1,0):~\{P_a,~J_{ab}\}.
    \end{align}
\end{subequations}
Similar to the Lorentzian class, we get either of the commutations
\begin{subequations}\label{e-commutations}
    \begin{align}
        &[\widetilde J_{AB},\widetilde{P}_C]=2\delta_{C[B}\widetilde{P}_{ A]},\quad [\widetilde J_{AB},\widetilde J_{CD}]=2\delta_{C[B}\widetilde J_{A]D}-2\delta_{D[B}\widetilde J_{A]C},\\
        &\text{or},\quad [\widetilde J_{ab},\widetilde{P}_c]=2\delta_{c[b}\widetilde{P}_{ a]},\quad [\widetilde J_{ab},\widetilde J_{cd}]=2\delta_{c[b}\widetilde J_{a]d}-2\delta_{d[b}\widetilde J_{a]c},
    \end{align}
\end{subequations}
respectively.\\
\newline
\underline{\textit{Summary}:} In summary, the unified flat geometry and algebra presented in \cref{unified-flat-geometry} and \eqref{unified-algbera}, derived via contraction of the Lorentzian flat geometry and Poincar\'e algebra, is significantly versatile. The inclusion of the parameter pair $(s,n)$ enables interpolation between multiple geometries and their associated isometry algebras. Notably, the commutation relations in \cref{n-commutations} remain almost inert under different interpretations, along with a small caveat. The choice of `$s$' triggers either of the changes: $\eta_{AB}\to\delta_{AB}$ or $\eta_{ab}\to \delta_{ab}$. Therefore, we need to take this into account while specializing to a specific case. Schematically speaking, it indicates that the covariant structure does not depend on $s$ or $n$, and may therefore be regarded, in a heuristic sense, as a covariantization of flat geometries.
\begin{quote}
    ``The upshot of this exercise is that any computation performed with unified structure, i.e., without fixing $(s,n)$, will yield a common result to all the cases.''
\end{quote}
\begin{figure}[h!]
    \centering
    \includegraphics[width=\linewidth]{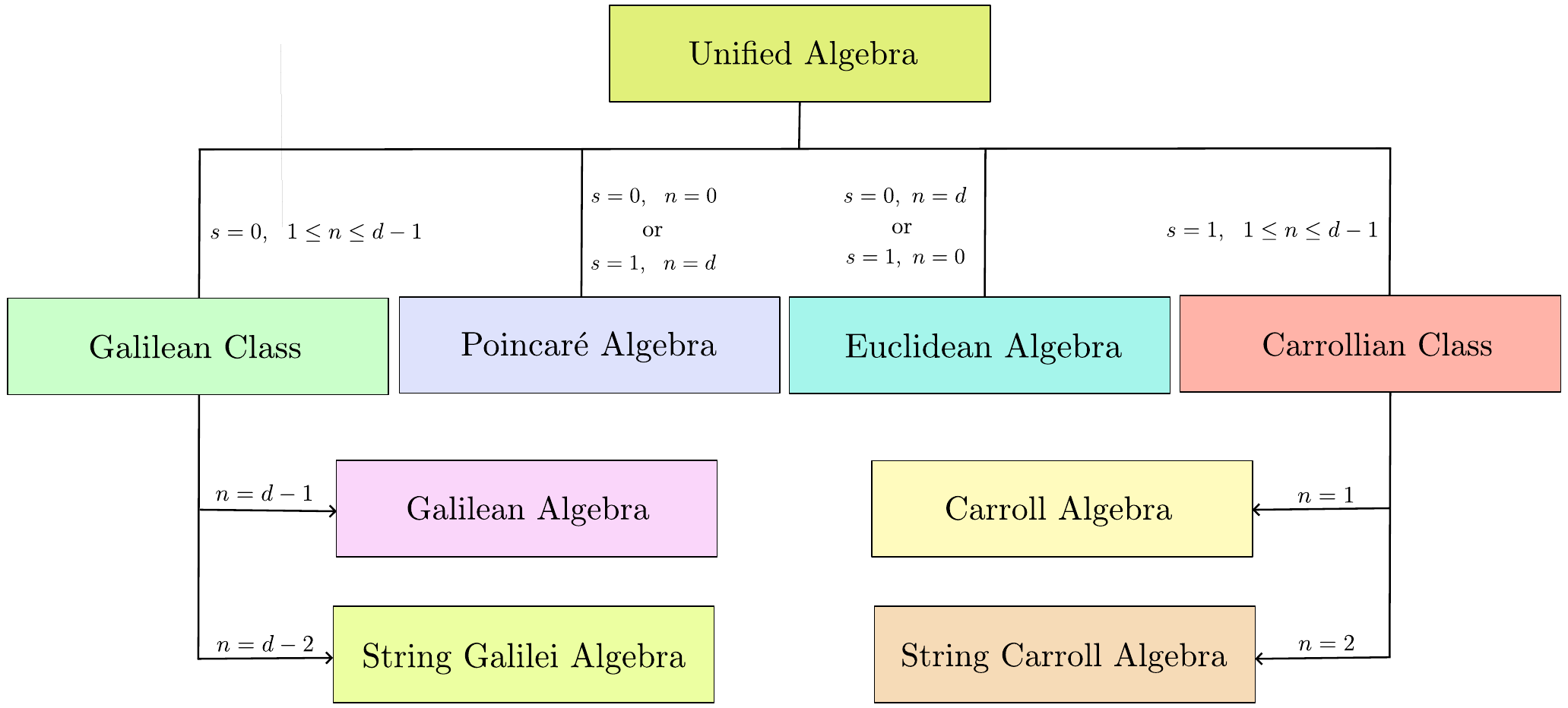}
    \caption{This is a schematic chart, which shows different choices of $s$ and $n$ corresponding to their respective symmetry algebra.}
    \label{fig:placeholder}
\end{figure}
This particular observation allows us to expand general relativity in a manner completely independent of any specific interpretation. In \cref{sec:GR-expansion}, we perform this expansion and subsequently, in \cref{sec:All-in one gravity}, we deduce its different realizations separately, which directly follow from this analysis.
\newpage
\section{Unified expansion of Lorentzian geometry}
\label{sec:Expansion-of-geometry}
As discussed in the previous section, a local metric on a spacetime manifold can be written in a unified form that admits different realizations upon fixing the pair $(s,n)$. In this section, we generalize this to a curved background by expanding Lorentzian geometry. We begin by choosing the local metric to take the form given in \cref{local-Lorentzian-flat}. Then we invoke an expansion of the frame fields, which we refer to as the \textit{unified expansion} (owing to the fact, $s$ and $n$ are left unfixed). This framework provides, in principle, a systematic way to retain subleading contributions and thereby capture deviations from the strict $\epsilon \to 0$ limit. However, in the present work, we restrict our attention to the leading-order data and its theory. We therefore leave the analysis of sub-leading contributions and the resulting theory for future work.

\medskip

First, let's consider the following ansatz for an expansion of vielbein and its inverse\footnote{From an algebraic perspective, these fields are interpreted as the gauge fields associated with local translations. And the gauge fields corresponding to local Lorentz transformations, namely the spin connections, can be written in terms of the vielbeins upon imposing curvature constraints that set the torsion to zero. However, this step is non-trivial, since the torsion-free condition may put constraints on geometry, refer to \cref{Torsion-zero-constraints}.} in orders of $\epsilon^2$. We expand them as 
\begin{subequations}\label{expansion-veilbeins}
    \begin{align}
        &\mathcal{E}^{\ A}_{\m}=e^{\ A}_{\m}+\mathcal{O}(\epsilon^2),\quad \mathcal{E}^{\ a}_{ \m}=e^{\ a}_{ \m}+\mathcal{O}(\epsilon^2),\\
        &\mathcal{E}_{\ A}^{ \m}=e_{\ A}^{ \m}+\mathcal{O}(\epsilon^2),\quad \mathcal{E}_{\ a}^{ \m}=e_{\ a}^{ \m}+\mathcal{O}(\epsilon^2).
    \end{align}
\end{subequations}
Before proceeding, it is important to emphasize the limitations of the even power expansion. Even though a background geometry can be described in different frames or charts, this choice must be made with care. In particular, an inappropriate choice of frame/chart may render the fields non-analytic in an even power expansion, thereby compromising the applicability of the subsequent analysis. We therefore assume the existence of a suitably well-behaved frame or coordinate system in which the expansion remains well defined. Indeed, one may also consider including odd powers in the expansion (refer to \cite{Ergen:2020yop} for an example of odd power expansion). However, their effects appear only in the sub-leading theory, which lies beyond the scope of the present work.

\medskip

We start by keeping the form of the local metric in \cref{local-Lorentzian-flat} intact and then write the spacetime metric accordingly to implement an expansion. As a consequence of the scale present in the local metric, we get the following decomposition of the spacetime metric and its inverse
\begin{equation}\label{Lorentzia-metric}
    g_{\mu\nu} = \epsilon^2 V_{\mu\nu} + \Pi_{\mu\nu}, 
    \qquad 
    g^{\mu\nu} = \frac{1}{\epsilon^2} V^{\mu\nu} + \Pi^{\mu\nu},
\end{equation}
where we have made the following redefinitions
\begin{equation}\label{def-in-frame-fields}
      V_{\mu\nu} = \eta_{AB}\,\mathcal{E}^{ \ A}_{ \mu}\,\mathcal{E}^{\ B}_{ \nu}, 
    \quad 
    V^{\mu\nu} = \eta^{AB}\,\mathcal{E}_{\ A}^{ \mu}\,\mathcal{E}_{\ B}^{ \nu},\quad \Pi_{\mu\nu} = \eta_{ab}\,\mathcal{E}^{\ a}_{ \mu}\,\mathcal{E}^{\ b}_{ \nu}, 
    \quad 
    \Pi^{\mu\nu} = \eta^{ab}\,\mathcal{E}_{\ a}^{ \mu}\,\mathcal{E}_{\ b}^{ \nu}.
\end{equation}
Therefore, we refer to $(V^{\m\n},V_{\m\n})$ as scaling metric data and to $(\Pi_{\m\n},\Pi^{\m\n})$ as non-scaling metric data. These redefined variables satisfy the following orthogonality and completeness relations,
\begin{equation}\label{SC-ortho-norm}
 V_{\m\n}\Pi^{\n\rho}=0,\quad V^{\m\n}\Pi_{\n\rho}=0, \quad V^{\m\s}V_{\s\n}+\Pi^{\m\s}\Pi_{\s\n}=\delta^{\m}_{\ \n}.
\end{equation}
Expansions in \cref{expansion-veilbeins} induces following expansion of the redefined variables
\begin{subequations}
    \begin{align}
    &V_{\m\n}=v_{\m\n}+\mathcal{O}(\epsilon^2),\quad V^{\m\n}=v^{\m\n}+\mathcal{O}(\epsilon^2),\\
        &\Pi_{\m\n}=h_{\m\n}+\mathcal{O}(\epsilon^2),\quad \Pi^{\m\n}=h^{\m\n}+\mathcal{O}(\epsilon^2),
    \end{align}
\end{subequations}
where the leading order variables are written in terms of the leading order frame fields as
\begin{subequations}\label{leading-order-variables}
    \begin{align}
         &v_{\mu\nu} = \eta_{AB}\,e^{\ A}_{ \mu}\,e^{\ B}_{ \nu}, 
    \qquad 
    v^{\mu\nu} = \eta^{AB}e_{ \ A}^{ \mu}e_{\ B}^{ \nu},\\
    &h_{\mu\nu} = \eta_{ab}e^{\ a}_{ \mu}e^{\ b}_{ \nu}, 
    \qquad 
    h^{\mu\nu} = \eta^{ab}e_{\ a}^{ \mu}e_{\ b}^{ \nu}.
    \end{align}
\end{subequations}
These fields encode the leading-order geometry, with the pair ($v^{\m\n}$, $h_{\m\n}$)\footnote{It naturally induces a fiber bundle structure on spacetime. The Lorentzian and Euclidean cases are exceptions because either the base or the fiber becomes zero-dimensional. Refer to Appendix~\ref{app:fibre bundle} for a discussion on this.} defining the unified geometry (we will see below, this is the only set of data invariant under local unified transformations). Furthermore, expanding the orthogonality and completeness relations in \cref{SC-ortho-norm} yields the constraints satisfied by the leading order geometry
\begin{equation}\label{LO-ortho-norm}
    v_{\mu\nu} h^{\nu\rho} = 0, 
    \quad 
    v^{\mu\nu} h_{\nu\rho} = 0, 
    \quad 
    v^{\mu\sigma}v_{\sigma\nu} + h^{\mu\sigma} h_{\sigma\nu} = \delta^{\mu}_{\ \nu}.
\end{equation}
We now show below that these leading-order variables indeed define a unified geometry. In particular, we now demonstrate that their local transformations (corresponding to unified symmetry) can be obtained from the expansion of local Lorentz transformations.\\
\newline
In general relativity, the Lorentzian vielbeins transform under the local Lorentz transformations: $\Lambda^A_{\ B}$, $\Lambda^A_{\ b}$ and $\Lambda^a_{\ b}$ as 
\begin{subequations}\label{local-lorentz-tranformations}
    \begin{align}
        &\delta_\Lambda \mathcal{E}^{\ A}_{ \m}=\Lambda^A_{\ B}\mathcal{E}^{\ B}_{ \m}+\Lambda^A_{\ b}\mathcal{E}^{\ b}_{ \m},\quad \delta_\Lambda \mathcal{E}^{\ a}_{ \m}=-\epsilon^2\eta_{AB}\eta^{ab}\Lambda^{ B}_{\ b}\mathcal{E}^{\ A}_{ \m}+\Lambda^a_{\ b}\mathcal{E}^{\ b}_{ \m},\\
        &\delta_\Lambda \mathcal{E}_{\ A}^{ \m}=-\Lambda_{\ A}^{ B}\mathcal{E}_{\ B}^{ \m}+\epsilon^2\eta_{AB}\eta^{ab}\Lambda^{B}_{\ a}\mathcal{E}_{\ b}^{ \m},\quad \delta_\Lambda \mathcal{E}_{\ a}^{ \m}=-\Lambda_{\ a}^{ B}\mathcal{E}_{\ B}^{ \m}-\Lambda_{\ a}^{ b}\mathcal{E}_{\ b}^{ \m}.
    \end{align}
\end{subequations}
It preserves the Minkowski metric $\tilde\eta$ on the frame bundle. Since the transformation parameter acts on vielbeins at every order of $\epsilon$, one should also assume an expansion of these parameters in order to maintain a well-defined perturbative hierarchy
\begin{equation}
    \Lambda^A_{\ B}=\lambda^A_{\ B}+\mathcal{O}(\epsilon^2),\quad \Lambda^A_{\ b}=\lambda^A_{\ b}+\mathcal{O}(\epsilon^2), \quad \Lambda^a_{\ b}=\lambda^a_{\ b}+\mathcal{O}(\epsilon^2).
\end{equation}
We now make a substitution of this expansion with \cref{expansion-veilbeins} in \cref{local-lorentz-tranformations}. At the leading order, we get the following variation of the geometry under local unified transformations
\begin{subequations}\label{local-transformation-unified}
    \begin{align}
        \delta_\lambda e^{\ A}_{ \m}&=\lambda^A_{\ B}e^{\ B}_{ \m}+\lambda^A_{\ b}e^{\ b}_{ \m},\\
        \delta_\lambda e^{\ a}_{ \m}&=\lambda^a_{\ b}e^{\ b}_{ \m},\\
        \delta_\lambda e_{\ A}^{ \m}&=-\lambda_{\ A}^{ B}e_{ \ B}^{ \m},\\
        \delta_\lambda e_{ \ a}^{ \m}&=-\lambda_{\ a}^{ B}e_{\ B}^{ \m}-\lambda_{\ a}^{ b}e_{\ b}^{ \m}.
    \end{align}
\end{subequations}
Using these transformations, it is a straightforward exercise to show that the leading order variables in \cref{leading-order-variables} transform as
\begin{equation}\label{local-unified-transformation}
    \delta_\lambda v_{\m\n}=2\eta_{AB}\lambda^A_{\ c}e^{\ B}_{ (\m}e^{\ c}_{\n)},\quad \delta_\lambda  h_{\m\n}=0,\quad \delta_\lambda v^{\m\n}=0,\quad \delta_\lambda h^{\m\n}=-2\eta^{ab}\lambda^B_{\ a}e_{\ B}^{ (\m}e_{\ b}^{\n)}.
\end{equation}
From this exercise, we have seen how local unified transformations arise from the $\epsilon \to 0$ limit of the local Lorentz transformations, which constitute a fundamental part of the emergent unified geometry. Although we can have certain tensors, such as $h^{\mu\nu}$ and $v_{\m\n}$, that are not inert under local transformations. But one can notice that certain combinations of these tensors that contract from local Lorentz invariant quantities ($\Phi$) remains invariant at the leading order
\begin{equation}
    \delta_\Lambda \Phi=0\xrightarrow{\epsilon\to0}\delta_\lambda \Phi^{(0)}=0,\quad \text{where}~\lim_{\epsilon\to0}\Phi=\Phi^{(0)}.
\end{equation}
Finally, we remark that the vielbeins of unified geometry and their transformation rules can also be obtained by a first principles approach, where we gauge the unified algebra \cref{unified-algbera}, refer to Appendix~\ref{app:Gauging}.

\section{A unified compatible connection}
\label{sec:Useful-connection}

As seen in the previous section, a unified geometry emerges from the expansion of a Lorentzian geometry. Formally, this geometry is defined by the pair $h_{\m\n}$ and $v^{\m\n}$ which is inert under local transformations. Therefore, it constitutes the intrinsic geometry data on a spacetime manifold.\\
\newline
Similar to general relativity we aim to compute a connection compatible with the unified geometry; we call it \textit{unified compatible connection}. As we will show in the next section, such a connection proves particularly useful in the expansion of the Einstein-Hilbert action, since it allows various expressions to be written in terms of unified covariant derivatives. In particular, the compatibility conditions permit free movement of $v^{\m\n}$ and $h_{\m\n}$ inside and out of the covariant derivative. Therefore such a connection satisfies following conditions
\begin{eqnarray}\label{SC-compatibility}
    \widetilde \N_\sigma v^{\m\n}=0,\quad \widetilde \N_\sigma h_{\m\n}=0.
\end{eqnarray} 
However, such a connection has no counterpart in the usual treatment of general relativity\footnote{Strictly speaking, this statement is not entirely true. As we will see later in \cref{sec:All-in one gravity}, in the Lorentzian and Euclidean class of the unified geometry, this connection reduces to the Levi-Civita connection. However, excluding these cases, the statement holds true.
}, where one works with the Levi-Civita connection satisfying the metric compatibility condition $\N_\sigma g_{\m\n}=0$. Therefore, the present connection cannot be obtained directly from the expansion of the Levi-Civita connection. One may instead impose a different compatibility condition on the Lorentzian counterpart, which in the limit $\epsilon\to0$ will reduce to \cref{SC-compatibility}. This then allows for a direct expansion of the connection, whose leading-order term yields a unified compatible connection. An analogous condition in the Lorentzian counterpart reads
\begin{eqnarray}\label{SC-compatibility-Lorentzian}
    \widehat \N_\sigma V^{\m\n}=0,\quad \widehat \N_\sigma \Pi_{\m\n}=0.
\end{eqnarray}
One such connection satisfying this condition is
\begin{eqnarray}\label{SC-connection-Lorentz}
    \widehat C^\rho_{\ \mu\nu}=V^{\rho\lambda}V_{\lambda\mu\nu}+\Pi^{\rho\alpha}\Pi_{\alpha\mu\nu}+\Pi^{\rho\alpha}V_{\nu\sigma}V^{\sigma\lambda}\Pi_{\lambda\mu\alpha},
\end{eqnarray}
where we have made the following redefinitions
\begin{subequations}
    \begin{align}
        V_{\lambda\mu\nu}&=\frac{1}{2}\left( \partial_\mu V_{\lambda\nu} + \partial_\nu V_{\mu\lambda} - \partial_\lambda V_{\mu\nu} \right),\\
       \Pi_{\lambda\mu\nu}&=\frac{1}{2}\left( \partial_\mu \Pi_{\lambda\nu} + \partial_\nu \Pi_{\mu\lambda} - \partial_\lambda \Pi_{\mu\nu} \right).
    \end{align}
\end{subequations}
A derivation of this connection is presented in Appendix \ref{app:derviation-of-connection}. In this appendix, we also show that the torsion in \cref{SC-connection-Lorentz} given by
\begin{eqnarray}
     \widehat T^\rho_{\mu\nu} = 2\widehat C^\rho_{[\mu\nu]}=2\Pi^{\rho\alpha}\Pi_{\lambda \alpha[\mu}V_{\nu]\sigma}V^{\sigma\lambda},
\end{eqnarray}
is intrinsic and stays finite in general (apart from the Lorentzian and Euclidean cases). Turning off this torsion imposes constraints on the geometry. 

\medskip

Now, as discussed above, we can obtain the unified compatible connection by taking $\epsilon\to0$ limit of $\widehat C^\rho_{\ \m\n}$ 
\begin{eqnarray}
    \lim_{\epsilon\to0} \widehat{C}^\rho_{\ \mu\nu}=\widetilde C^\rho_{\ \mu\nu}=v^{\rho\lambda}v_{\lambda\mu\nu}+h^{\rho\alpha}(h_{\alpha\mu\nu}+v_{\nu\sigma}v^{\sigma\lambda}h_{\lambda\mu\alpha}),
\end{eqnarray}
where,
\begin{subequations}\label{notation-1}
    \begin{align}
        &v_{\lambda\mu\nu}=\frac{1}{2}\left( \partial_\mu v_{\lambda\nu} + \partial_\nu v_{\mu\lambda} - \partial_\lambda v_{\mu\nu} \right),\\
        &h_{\lambda\mu\nu}=\frac{1}{2}\left( \partial_\mu h_{\lambda\nu} + \partial_\nu h_{\mu\lambda} - \partial_\lambda h_{\mu\nu} \right).
    \end{align}
\end{subequations}
Similarly, we can compute its torsion
\begin{eqnarray}
     \lim_{\epsilon\to0}\widehat T^\rho_{\mu\nu} = \widetilde T^\rho_{\mu\nu} =2h^{\rho\alpha}h_{\lambda \alpha[\mu}v_{\nu]\sigma}v^{\sigma\lambda}.
\end{eqnarray}
We further quote some useful identities that the connection \cref{SC-compatibility-Lorentzian} satisfies
\begin{eqnarray}
    \widehat \nabla_\sigma V_{\mu\nu}=2V_{\rho\sigma(\mu}\Pi_{\ \nu)\lambda}\Pi^{\lambda\rho},\quad \widehat \nabla_\sigma \Pi^{\mu\nu}=- 2 \Pi^{\lambda(\mu} V^{\nu)\rho}\widehat \nabla_\sigma V_{\lambda\rho}.
\end{eqnarray}
Furthermore, a useful quantity associated with this connection is its trace, which is given by
\begin{eqnarray}
    \widehat C^\rho_{\ \rho \n}=\frac{1}{E}\del_\n E+\Pi^{\rho\alpha}V_{\nu\sigma}V^{\sigma\lambda}\Pi_{\lambda\rho\alpha},
\end{eqnarray}
where $E$ is defined via the Jacobian $\sqrt{-g}=\epsilon^n E=\epsilon^n\sqrt{-\det(V_{\m\n}+\Pi_{\m\n})}$. A detailed derivation for the $n=2$ case is presented in Appendix-E of ref.~\cite{Bagchi:2026qpi}; this is a straightforward generalization. This result enables us to express the divergence involving this covariant derivative as a total derivative, together with an additional term
\begin{eqnarray}
    \widehat \nabla_\mu X^\mu=\del_\mu X^\mu +\widehat C^\rho_{\ \rho\nu}X^\nu=\frac{1}{E}\del_\mu(E X^\mu)+\Pi^{\rho\alpha}V_{\nu\sigma}V^{\sigma\lambda}\Pi_{\lambda\rho\alpha}  X^\nu
\end{eqnarray}
Further, it allows to make the following simplification
\begin{eqnarray}
    \int_\mathcal{M}\widehat \nabla_\mu X^\mu ~E~d^{d}x\approx\int_\mathcal{M} (\Pi^{\rho\alpha}V_{\nu\sigma}V^{\sigma\lambda}\Pi_{\lambda\rho\alpha}  X^\nu)~E~d^{d}x.
\end{eqnarray}
Where $\approx$ denotes the value of the integral after dropping a boundary term. The boundary term will become relevant if $\mathcal{M}$ has a finite boundary or the field has non-trivial fall-offs.
\section{Unified decomposition of Einstein-Hilbert action}
\label{sec:GR-expansion}
We now have all the necessary ingredients to systematically perform an expansion of the Einstein–Hilbert action. To this end, we adopt the following strategy. We begin with the Levi-Civita connection and, instead of expanding it directly, organize it in terms of the Lorentzian variables, arranged by the expansion parameter $\epsilon$. We call this a \textit{unified decomposition}. In the same spirit, we arrange the curvature tensors and the Ricci scalar accordingly. This procedure allows us to rewrite the Lorentzian theory in a form suitable for extracting contributions at arbitrary orders in the expansion. However, we skip the expansion in this section and discuss it in \cref{sec:All-in one gravity} after specializing the decomposition obtained here to particular cases.
\\
\newline
\underline{\textit{Levi-Civita Connection}:} We now perform a unified decomposition of the Levi-Civita connection, showing that the affine connection \cref{SC-connection-Lorentz} can be incorporated into the decomposition by adding and subtracting the torsional contribution.\\
\newline
The decomposed Levi-Civita connection reads
\begin{align}\label{SC-decomposition-Levi-Civtia}
    \Gamma^\rho_{\ \mu\nu}&=\frac{g^{\rho\lambda}}{2}\left(\del_\mu g_{\nu\lambda}+\del_\nu g_{\mu\lambda}-\del_\lambda g_{\mu\nu}\right)\notag\\
        &=\frac{1}{\epsilon^2}\overset{(-2)}{\Gamma^\rho_{\ \mu\nu}}+\widehat C^\rho_{\ \mu\nu}+\mathcal{T}^\rho_{\mu\nu}+\epsilon^2\overset{(2)}{\Gamma^\rho_{\mu\nu}},
\end{align}
where we have introduced a shift tensor $\mathcal{T}^\rho_{\ \mu\nu}$ following a prescription similar to that used in the pre-ultra-local parametrization \cite{Hansen:2021fxi}. This shift tensor contains the torsional part of $\widehat C^\rho_{\ \mu\nu}$ with an opposite sign. All terms in the decomposition are given by
\begin{subequations}
    \begin{align}
      \overset{(-2)}{\Gamma^\rho_{\ \mu\nu}}& =V^{\rho\lambda}\Pi_{\lambda\m\n},\\
      \mathcal{T}^\rho_{\mu\nu}&=-\Pi^{\rho\alpha }V_{\n\sigma}V^{\sigma\lambda}\Pi_{\lambda\m \alpha},\\
      \overset{(2)}{\Gamma^\rho_{\mu\nu}}&=\Pi^{\rho\lambda}V_{\lambda\m\n}.
    \end{align}
\end{subequations}
Using this decomposition, we can now proceed with the unified decomposition of the Ricci tensor and scalar. Before that, let's write some useful relations:
\begin{align}
        \overset{(-2)}{\Gamma^\rho_{\ \rho\nu}}=0,\quad  \overset{(2)}{\Gamma^\rho_{\ \rho\nu}}=0,\quad \mathcal{T}^\rho_{\mu\nu} V^{\m\n}=0,\quad \mathcal{T}^\rho_{\mu\nu} \Pi^{\n\sigma}=0,\quad V^{\m\n} \overset{(-2)}{\Gamma^\rho_{\ \mu\nu}}=0,\quad \Pi^{\m\n}\overset{(2)}{\Gamma^\rho_{\ \mu\nu}}=0.
\end{align}
\underline{\textit{Ricci tensor and Ricci scalar}:}  The Ricci tensor of the
Levi-Civita connection is decomposed in terms of Lorentzian variables using \cref{SC-decomposition-Levi-Civtia} as
\begin{subequations}
    \begin{align}
        R_{\mu\nu}&=-\del_\mu \Gamma^\rho_{\ \rho\nu}+\del_\rho \Gamma^\rho_{\ \mu\nu}-\Gamma^\rho_{\ \mu\lambda}\Gamma^\lambda_{\ \rho\nu}+\Gamma^\rho_{\ \rho\lambda}\Gamma^\lambda_{\ \mu\nu}\\
        &=\frac{1}{\epsilon^4}\overset{(-4)}{R_{\mu\nu}}+\frac{1}{\epsilon^2}\overset{(-2)}{R_{\mu\nu}}+\overset{(0)}{R_{\mu\nu}}+\epsilon^2\overset{(2)}{R_{\mu\nu}}+\epsilon^4\overset{(4)}{R_{\mu\nu}}.
    \end{align}
\end{subequations}
The terms in the decomposition are given by
\begin{subequations}
    \begin{align}
        &\overset{(-4)}{R_{\mu\nu}}=-\overset{(-2)}{\Gamma^\rho_{\ \mu\lambda}}\overset{(-2)}{\Gamma^\lambda_{\ \rho\nu}},\\
        &\overset{(-2)}{R_{\mu\nu}}=\widehat{\nabla}_\rho \overset{(-2)}{\Gamma^\rho_{\ \mu\nu}}-\mathcal{T}^\rho_{\ \mu\lambda}\overset{(-2)}{\Gamma^\lambda_{\ \rho\nu}}+\mathcal{T}^\rho_{\ \rho\lambda}\overset{(-2)}{\Gamma^\lambda_{\ \mu\nu}}-2\widehat{C}^\rho_{\ [\mu\lambda]}\overset{(-2)}{\Gamma^\lambda_{\ \rho\nu}}-\overset{(-2)}{\Gamma^\rho_{\ \mu\lambda}}\mathcal{T}^\lambda_{\ \rho\nu},\\
        &\overset{(0)}{R_{\mu\nu}}=\widehat{R}_{\mu\nu}-\widehat{\nabla}_\mu \mathcal{T}^\rho_{\ \rho\nu}+\widehat{\nabla}_\rho \mathcal{T}^\rho_{\ \mu\nu}-\overset{(-2)}{\Gamma^\rho_{\ \mu\lambda}}\overset{(2)}{\Gamma^\lambda_{\ \rho\nu}}-\overset{(2)}{\Gamma^\rho_{\ \mu\lambda}}\overset{(-2)}{\Gamma^\lambda_{\ \rho\nu}}-2 \widehat{C}^\lambda_{\ [\mu\rho]}\mathcal{T}^\rho_{\ \lambda\nu},\\
        &\overset{(2)}{R_{\mu\nu}}=\widehat{\nabla}_\rho \overset{(2)}{\Gamma^\rho_{\ \mu\nu}}-2 \widehat{C}^\rho_{ [\mu\lambda] }\overset{(2)}{\Gamma^\lambda_{\ \rho\nu}}-\mathcal{T}^\lambda_{\ \rho\nu}\overset{(2)}{\Gamma^\rho_{\ \mu\lambda}},\\
        &\overset{(4)}{R_{\mu\nu}}=- \overset{(2)}{\Gamma^\rho_{\ \mu\lambda}}\overset{(2)}{\Gamma^\lambda_{\ \rho\nu}}.
    \end{align}
\end{subequations}
where we have defined $\widehat R_{\mu\nu}$ as the Ricci tensor for the connection $\widehat C^\rho_{\ \mu\nu}$. Now we could perform the unified decomposition of the Ricci scalar in a similar way
\begin{eqnarray}\label{Ricci-decomp}
    R&&=g^{\mu\nu}R_{\mu\nu}=\left(\frac{1}{\epsilon^2}V^{\mu\nu}+\Pi^{\mu\nu}\right)\left(\frac{1}{\epsilon^4} \overset{(-4)}{R}_{\mu\nu} +
\frac{1}{\epsilon^2} \overset{(-2)}{R}_{\mu\nu} +
\overset{(0)}{R}_{\mu\nu} +
\epsilon^2\, \overset{(2)}{R}_{\mu\nu} +
\epsilon^4\, \overset{(4)}{R}_{\mu\nu}\right)\notag\\
&&\nonumber=\frac{1}{\epsilon^6}V^{\m\n}\overset{(-4)}{R}_{\mu\nu}+\frac{1}{\epsilon^4}\left(\Pi^{\mu\nu}\overset{(-4)}{R}_{\mu\nu}+V^{\mu\nu}\overset{(-2)}{R}_{\mu\nu}\right)+\frac{1}{\epsilon^2}\left(\Pi^{\mu\nu}\overset{(-2)}{R}_{\mu\nu}+V^{\mu\nu}\overset{(0)}{R}_{\mu\nu}\right)\\
&&\hspace{1cm}+\left(\Pi^{\mu\nu}\overset{(0)}{R}_{\mu\nu}+V^{\mu\nu}\overset{(2)}{R}_{\mu\nu}\right)+\epsilon^2\left(\Pi^{\mu\nu}\overset{(2)}{R}_{\mu\nu}+V^{\mu\nu}\overset{(4)}{R}_{\mu\nu}\right)+\epsilon^4~\Pi^{\mu\nu}\overset{(4)}{R}_{\mu\nu}.
\end{eqnarray}
It can be shown $V^{\mu\nu}\overset{(-4)}{R}_{\mu\nu}=0$, $\Pi^{\mu\nu}\overset{(4)}{R}_{\mu\nu}=0$ and other non-trivial terms are
\begin{subequations}
    \begin{align}
         \Pi^{\mu\nu}\overset{(-4)}{R}_{\mu\nu}+V^{\mu\nu}\overset{(-2)}{R}_{\mu\nu} &=\Pi^{\rho\alpha}V^{\beta\kappa}V^{\m\n}\Pi_{\kappa\rho\n}\Pi_{\m\alpha\beta},\\\Pi^{\mu\nu}\overset{(-2)}{R}_{\mu\nu}+V^{\mu\nu}\overset{(0)}{R}_{\mu\nu}&=2\widehat\N_\m X^\m -2V^{\sigma \lambda}\Pi_{\sigma \beta\m}\Pi_{\lambda \rho \n}\Pi^{\m(\n}\Pi^{\beta)\rho}+V^{\m\n}\widehat R_{\m\n},\\
         \Pi^{\mu\nu}\overset{(0)}{R}_{\mu\nu}+V^{\mu\nu}\overset{(2)}{R}_{\mu\nu}&=\widehat\N_\m Y^\m +\Pi^{\m\n}\widehat R_{\m\n},\\
         \Pi^{\mu\nu}\overset{(2)}{R}_{\mu\nu}+V^{\mu\nu}\overset{(4)}{R}_{\mu\nu}&=\Pi^{\rho\beta}V_{\beta\m\n}V^{\m \alpha}\Pi^{\n \kappa}V_{\kappa \rho \alpha},
    \end{align}
\end{subequations}
where we have defined $X^\m=V^{\m\lambda}\Pi_{\lambda\alpha\beta}\Pi^{\alpha\beta}$ and $Y^\m=\Pi^{\m\lambda}V_{\lambda\alpha\beta}V^{\alpha\beta}$. With all the necessary ingredients at hand, we can now construct a systematic and covariant unified decomposition of general relativity.\\
\newline
\underline{\textit{Einstein-Hilbert Action}:} We start with the Einstein-Hilbert action with cosmological constant $\Lambda$
\begin{eqnarray}\label{EH-Action}
    S=\frac{1}{16 \pi G}\int d^d x~\sqrt{-g}~(R-2\Lambda),
\end{eqnarray}
where, $\sqrt{-g}=\epsilon^n E$ as defined previously.

\medskip 

However, the cosmological constant does not come equipped with a canonical prescription for its expansion in powers of $\epsilon$\footnote{Expanding the cosmological constant may not admit a direct physical interpretation for a given choice of $s$ and $n$. However, we keep it for the sake of generality.} \footnote{Equivalently, the gravitational constant $G$ can also be expanded, and an expansion hierarchy can be maintained. However, keeping track of such an expansion is not a primary scope of the present work, we refrain from considering it here.}. The precise way of its expansion is therefore a matter of choice and is determined by the particular gravitational theory one wishes to obtain at leading order. Depending on this choice, the resulting leading-order theory may or may not contain a cosmological constant term. Also, this choice does not spoil the symmetry of the leading-order theory, because it enters the action only as a scalar contribution. With this in mind, we consider the following general expansion for the cosmological constant 
\begin{eqnarray}
    \Lambda=\epsilon^\Delta \widehat \Lambda=\epsilon^\Delta\left(\overset{\{\Delta\}}{\Lambda}+\overset{\{\Delta+2\}}{\Lambda}\epsilon^2+\overset{\{\Delta+4\}}{\Lambda}\epsilon^4+\cdots\right).
\end{eqnarray}
The parameter $\Delta$ will be fixed once a specific choice of the pair ($s,n$) is made and the power of $\epsilon$ of the leading-order theory is known. 

\medskip

Let's now invoke the unified decomposition of the Ricci scalar \cref{Ricci-decomp} in \cref{EH-Action}. We get the following decomposition of the Einstein-Hilbert action
\begin{align}\label{unified-EH}
     S\approx \frac{1}{16\pi G}\int d^dx ~E~ \Big[\epsilon^{n-4} L^{(n-4)}+\epsilon^{n-2} L^{(n-2)}
    +\epsilon^{n} L^{(n)}+\epsilon^{n+2} L^{(n+2)}-2\epsilon^{\Delta+n} \widehat\Lambda\Big],
\end{align}
where,
\begin{subequations}
    \begin{align}
       \label{LO-decomp-EH} L^{(n-4)}&=\Pi^{\rho\alpha}V^{\beta\kappa}V^{\m\n}\Pi_{\kappa\rho\n}\Pi_{\m\alpha\beta},\\
         L^{(n-2)}&=2V^{\sigma \lambda}\Pi_{\sigma \beta\m}\Pi_{\lambda \rho \n}\Pi^{\m[\beta}\Pi^{\nu]\rho}+V^{\m\n}\widehat R_{\m\n},\\
          L^{(n)}&=\Pi^{\m\n}\widehat R_{\m\n},\\
           L^{(n+2)}&=\Pi^{\rho\beta}V_{\beta\m\n}V^{\m \alpha}\Pi^{\n \kappa}V_{\kappa \rho \alpha}.
    \end{align}
\end{subequations}
This decomposition is one of the main results of this paper. As emphasized earlier, the unified decomposition remains valid for all possible choices of the pair $(s,n)$. However, we do not take the $\epsilon\to0$ limit immediately. In the next section, we will see that, for certain cases, the leading-order term in the decomposition \cref{LO-decomp-EH} vanishes identically. Consequently, in such cases, it does not contribute to the leading-order term in the expansion. Though it is not an issue, we conclude this section here to avoid confusion.
\section{Avatars of unified decomposition and expansion}
\label{sec:All-in one gravity}
Till now, we have computed a unified decomposition of the Einstein-Hilbert action with a finite cosmological constant. In this section, we borrow the identifications discussed in \cref{sec:Symmetries} to specialize the unified decomposition of the gravity action. Subsequently, we map the leading order theory to general relativity\footnote{Though this case seems a bit confusing at first glance. The expansion of Einstein's gravity appears to reproduce Einstein's gravity itself as the leading-order theory. One may therefore wonder what non-trivial information is encoded in the expansion. In the subsequent \cref{subec:GR}, we see a non-trivial example for this.}, non-relativistic gravity, and Carroll gravity. However, for the specialization and a detailed discussion on string Carroll gravity, refer to \cref{subsec: String Carroll gravity}.

\subsection{General relativity}
\label{subec:GR}
In this part of the section, we show how general relativity is encoded within the unified decomposition of the Einstein-Hilbert action. Although the identification is straightforward in this case, we present it because the expansion applies to a non-trivial example of near-horizon expansion of extremal BHs via KLR prescription.

\medskip

As discussed in \cref{sec:Symmetries}, the Poincar\'e algebra arises for two distinct choices of the pair $(s,n)$\footnote{The discussion for the Euclidean class proceeds analogously. Therefore, we omit it for brevity.}. We now show how these choices are reflected in the redefined variables introduced in \cref{def-in-frame-fields}.\\
\newline 
$\underline{(s,n)\to(0,0):}$ In this case, we only retain the lowercase Latin indices. Consequently, this choice implies
\begin{eqnarray}
    V_{\mu\nu}=0,\qquad 
    V^{\mu\nu}=0.
\end{eqnarray}

Let us now examine the effect of this choice on the orthogonality and completeness relation \cref{SC-ortho-norm}. The only non-trivial condition remaining is
\begin{eqnarray}
    \Pi^{\mu\sigma}\Pi_{\sigma \nu}=\delta^\mu_{\ \nu}.
\end{eqnarray}
This relation implies that $\Pi^{\mu\nu}$ is the inverse metric of $\Pi_{\mu\nu}$\footnote{For the Galilean and Carrollian classes, this notion no longer exists. Since $\Pi_{\mu\nu}$ becomes degenerate and therefore admits no inverse.}. Also, setting $s=0$ makes $\Pi_{\m\n}$ carry one negative eigenvalue, which makes it a Lorentzian metric. Consequently, the affine connection in \cref{SC-connection-Lorentz} reduces to the Levi-Civita connection
\begin{eqnarray}
    \widehat C^\rho_{\ \mu\nu}
    =
    \Pi^{\rho\beta}\Pi_{\beta\mu\nu}.
\end{eqnarray}
As expected, the torsional contribution drops out. Therefore, the Ricci tensor $\widehat R_{\mu\nu}$ reduces to the usual Ricci tensor constructed from the Levi-Civita connection.

The unified decomposition of the Einstein-Hilbert action in \cref{unified-EH} then becomes
\begin{eqnarray}
    S
    =
    \frac{1}{16\pi G}
    \int_\mathcal{M} d^d x~E~
    \left(
    \Pi^{\mu\nu}\widehat R_{\mu\nu}
    -
    2\widehat\Lambda~\epsilon^\Delta
    \right)
    =
    \frac{1}{16\pi G}
    \int_\mathcal{M} d^d x~E~
    \left(
    \widehat R
    -
    2\widehat\Lambda~\epsilon^\Delta
    \right),
\end{eqnarray}
where $E=\sqrt{-\det \Pi_{\m\n}}$. Now, choosing $\Delta=0$ yields general relativity with a finite cosmological constant in the limit $\epsilon\to0$. On the other hand, any choice with $\Delta>0$ suppresses the cosmological constant in this limit\footnote{\label{footnote-9}The case with $\Delta<0$ is rather trivial. In the limit $\epsilon\to 0$, we only get
\begin{eqnarray}
    \lim_{\epsilon\to0}S=-\frac{1}{8\pi G}\int_\mathcal{M}d^dx ~e~\overset{\{\Delta \}}{\Lambda}.
\end{eqnarray}
Consequently, the equation of motion for the leading-order metric simply sets
\begin{eqnarray}
    \overset{\{\Delta\}}{\Lambda}=0.
\end{eqnarray}
The same argument applies iteratively to all terms of order $\epsilon^r$ with $r<0$ and first non-triviality appears at $r=0$, which is similar to the case $\Delta=0$.}. The main point is that the standard Einstein's theory of gravity is recovered directly from the unified decomposition. Therefore, in the limit $\epsilon\to0$ we get
\begin{eqnarray}\label{eq: lo eh action non scaling}
    \lim_{\epsilon\to0}S=\frac{1}{16\pi G} \int_\mathcal{M} d^dx~e \mathcal L_{LO} \,,
\end{eqnarray}
where the Jacobian is defined as $e=\sqrt{-\det h_{\m\n}}$ and
\begin{equation}
    \mathcal L_{LO} = \begin{cases}
       \widetilde R-2\overset{\{0\}}{\Lambda},&\Delta=0\\
         \widetilde R,&\Delta>0
    \end{cases} \;.
\end{equation}

At the leading order we get Einstein's gravity. But we have yet to specify the physical meaning of the expansion and the expansion parameter. We will briefly discuss this later after presenting the second realization of the Lorentzian case.\\
\newline
$\underline{{(s,n)\to(1,d):}}$ In contrast to the previous case, we are left with the capital Latin indices, which consequently sets
\begin{eqnarray}
    \Pi_{\m\n}=0,\quad \Pi^{\m\n}=0.
\end{eqnarray}
With this choice, only the following non-trivial orthogonality and completeness relation remains
\begin{eqnarray}
    V^{\m\sigma}V_{\sigma \n}=\delta^\m_{\ \n}.
\end{eqnarray}
It says that $V^{\m\n}$ is the inverse of $V_{\m\n}$. Since we have $s=1$, they form geometric data for a Lorentzian geometry. As a result, the affine connection in \cref{SC-connection-Lorentz} just becomes 
\begin{eqnarray}\label{lorentzian-case-2-met}
    \widehat{C}^\rho_{\ \m\n}=V^{\rho\sigma}V_{\sigma \m\n},
\end{eqnarray}
which is again the usual Levi-Civita connection. Therefore, the Ricci tensor $\widehat R_{\m\n}$ reduces to the Ricci tensor for the Levi-Civita connection \cref{lorentzian-case-2-met}. Now the unified decomposition of the Einstein-Hilbert action becomes
\begin{eqnarray}
    S=\frac{1}{16\pi G}\int_{\mathcal{M}}d^d x~E~\left(\epsilon^{d-2}V^{\m\n}\widehat R_{\m\n}-2\epsilon^{d+\Delta}\widehat \Lambda\right),
\end{eqnarray}
where, $E=\sqrt{-\det V_{\m\n}}$. Again, we can make different choices for $\Delta$ to obtain two distinct classes of theory at the leading order in expansion. The first possibility is $\Delta=-2$, which leads to Einstein's gravity with a finite cosmological constant at the leading order in the expansion. Second, is $\Delta>-2$, where the leading order theory has a vanishing cosmological constant. The remaining case with $\Delta<-2$ is again trivial, as it follows the same reasoning of $\Delta<0$ case for $(s,n)=(0,0)$ (see \cref{footnote-9}). This choice effectively reduces to the $\Delta=-2$ case. Therefore in the limit $\epsilon\to0$ we get
\begin{eqnarray}
    \lim_{\epsilon\to0}\epsilon^{2-d}S=\frac{1}{16\pi G} \int_\mathcal{M} d^dx~e\begin{cases}
       \widetilde R-2\overset{\{-2\}}{\Lambda},&\Delta=-2,\\
         \widetilde R,&\Delta>-2,
    \end{cases}
\end{eqnarray}
where, the Jacobian is given by $e=\sqrt{-\det v_{\m\n}}$. We have now seen both cases in which general relativity emerges from the unified expansion at leading order. However, the physical meaning of the expansion itself still remains to be understood.\\
\newline
\underline{\textit{Comment on the expansion}:} In the above discussed cases, leading-order geometry is a solution of Einstein's gravity, whereas the full geometry (unexpanded version) also satisfies Einstein's field equations. Then an immediate question arises: what is the physical meaning of this expansion? 

\medskip

Roughly speaking, corrections arising from this expansion may be regarded as dynamical fluctuations of a Lorentzian geometry that satisfies the Einstein field equations, where fluctuations are also Lorentzian. Apart from this, there is a very interesting example of such an expansion: a near-horizon expansion of extremal BHs. A well-known universal spacetime structure of the extremal BHs is known to appear in the near-horizon region, which develops an $AdS_2$ throat. Another important feature is that the leading-order geometry obtained in the near-horizon expansion satisfies the Einstein equations. This feature is exactly the same as the Lorentzian case of the unified expansion. Let's see this with an explicit example of an extremal Kerr BH.
\subsubsection{Near-horizon of extremal Kerr BH via KLR prescription}
The near-horizon geometry of the extremal Kerr black hole, obtained via the Kunduri-Lucietti-Reall (KLR) prescription \cite{Kunduri:2007vf}, admits an $AdS_2$ throat\footnote{We refer readers to the following review article for more details on this topic: \cite{Kunduri:2013gce}.}. This example describes ${{{(s,n)\to(0,0)}}}$ case of unified expansion, with $\widehat{\Lambda}=0$ (since extremal Kerr satisfies vacuum Einstein equation). The extremal Kerr metric in Boyer-Lindquist coordinates is
\begin{equation}
    g = - \frac{\widetilde\Delta}{\Sigma} \left( dt - a \sin^2 \theta \, d\phi \right)^2 + \frac{\sin^2 \theta}{\Sigma} \left( a dt - (r^2 + a^2 ) d\phi \right)^2 + \frac{\Sigma}{\widetilde\Delta} dr^2 + \Sigma \, d\theta^2 \,,
\end{equation}

where,
\begin{equation}
    \Sigma = r^2 + a^2 \cos^2\theta \,, \qquad \widetilde\Delta = (r-a)^2\,, \qquad a =\frac{J}{M}  \,.
\end{equation}

The event horizon sits at the radial location $r = a$, where $\tilde\Delta$ vanishes. Now we make the following coordinate transformation before proceeding to near-horizon expansion: $(t,r,\theta,\phi) \to (\tilde t, \rho, \theta, \tilde\phi)$, defined as
\begin{gather}
    r = a + 2\varepsilon\rho \,, \quad t = \frac{\tilde t}{\varepsilon} \,, \quad \phi = \tilde\phi + \frac{\tilde t }{2 a \varepsilon} .
\end{gather}

The Kerr metric thus expands as
\begin{equation}\label{eq: nh extremal kerr}
    g = h_{\m\n}dx^\m dx^\n + \mathcal{O}(\varepsilon) \,,
\end{equation}
where the limit $\varepsilon\to 0$ gives us the near-horizon extremal Kerr geometry
\begin{equation}
    h_{\m\n}dx^\m dx^\n =  (1+\cos^2 \theta) \left[ \underbrace{-\frac{\rho^2}{a^2}d\tilde t^{\,2}  + \frac{a^2}{\rho^2}d\rho^2}_{AdS_2\, \text{ throat}} + a^2 d\theta^2 \right] + \frac{4 a^2 \sin^2\theta}{1 + \cos^2 \theta}\left( \frac{\rho}{a^2} d\tilde t + d\tilde\phi \right)^2 \, \,.
\end{equation}

One can check that the leading-order metric satisfies Einstein's field equations \cite{Bardeen:1999px}. Further, we can map $\varepsilon$ with the unified expansion parameter $\epsilon$ as
\begin{eqnarray}
    \varepsilon=\epsilon^2.
\end{eqnarray}
Therefore, we can conclude that the near-horizon expansion of the extremal Kerr BH in 4$D$ using the KLR prescription is an example of the Lorentzian case of the unified expansion of Einstein's gravity. 

\subsection{Non-relativistic gravity}
\label{subsec: non-relativistic}
Now we choose $s$ and $n$, such that the unified local symmetries reduce to the Galilean one. As shown in \cref{sec:Symmetries}, this corresponds to the following choice of these parameters
\begin{eqnarray}
    s=0,\quad n=d-1
\end{eqnarray}
This enforces capital Latin indices to run over $d-1$ components, whereas lowercase Latin indices run over a single time-like index. As a result, the redefined variables in \cref{def-in-frame-fields} take the following form
\begin{eqnarray}\label{galilean-vielbeins}
    V_{\m\n}=\delta_{AB}\mathcal{E}_\m^{\ A}\mathcal{E}_\n^{\ B},\quad V^{\m\n}=\delta^{AB} \mathcal{E}^\m_{\ A}\mathcal{E}_{\ B}^{\n},\quad \Pi_{\m\n}=-\Pi_\m \Pi_\n,\quad  \Pi^{\m\n}=-\Pi^\m \Pi^\n
.\end{eqnarray}
where we have represented $\mathcal{E}_\m^{\ t}$ and $\mathcal{E}^\m_{\ t}$ with $\Pi_\m$ and $-\Pi^\m$ respectively. In the limit $\epsilon\to 0$ they reduce to 
\begin{eqnarray}
    \lim_{\epsilon\to0}\Pi_\m=h_{\m},\quad \lim_{\epsilon\to0}\Pi^\m=h^\m.
\end{eqnarray}
It is a simple check from \cref{local-unified-transformation} that the local Galilean invariant data is given by
\begin{eqnarray}
    \delta_\lambda v^{\m\n}=0,\quad \delta_\lambda h_\m=0.
\end{eqnarray}
which is consistent with the transformations of geometry on a Newton-Cartan manifold \cite{Duval:2014uoa} (equivalently, we call it a Galilean geometry), with $h_{\m}$ being the invariant clock 1-form. Also, the orthogonality and normalization relations become
\begin{eqnarray}\label{eq: galilean data}
   h^\m h_\m=-1 ,\quad v_{\m\n}h^\n=0,\quad v^{\m\n}h_\n=0,\quad -h^\m h_\n+v^{\m\s}v_{\s\n}=\delta^\m_{\ \n}.
\end{eqnarray}
Further, this choice reduces the affine connection in \cref{SC-connection-Lorentz} to \footnote{Refer to Appendix \ref{app:mapping connection} for details on this mapping.}
\begin{eqnarray}\label{PNR-connection}
    \widehat{C}^\rho_{\ \m\n}=-\Pi^\rho \del_\m \Pi_\n+V^{\rho \lambda}V_{\lambda \m\n}.
\end{eqnarray}
This is exactly the pre-non-relativistic connection defined in \cite{c}\footnote{It carries a finite torsion given by
\begin{eqnarray}
    \widehat{T}^\rho_{\ \m\n}=2\widehat C^\rho_{\ [\m\n]}=-2\Pi^\rho \del_{[\m}\Pi_{\n]}.
\end{eqnarray}}. Now we proceed to evaluate the unified decomposition of the Einstein-Hilbert action under the above considerations. The decomposition in \cref{unified-EH} then becomes
\begin{align}
     S\approx \frac{1}{16\pi G}\int d^dx ~E~ \Big[&\epsilon^{d-5}\left(V^{\m\n}V^{\beta\kappa}\del_{[\kappa}\Pi_{\n]}\del_{[\beta}\Pi_{\m]}\right)+\epsilon^{d-3}V^{\m\n}\widehat R_{\m\n}\notag \\&\quad ~~~~~~~~~~~~~-\epsilon^{d-1}\Pi^{\m}\Pi^{\n}\widehat R_{\m\n}-2\epsilon^{\Delta+d-1} \widehat\Lambda\Big],
\end{align}
where, $E=\sqrt{-\det(V_{\m\n}-\Pi_\m \Pi_\n)}$. This maps to the pre-non-relativistic decomposition of general relativity, with an extension to include a cosmological constant. Here as well, we can make various choices of $\Delta$ to obtain different theories at leading order.

\medskip

For $\Delta=-4$ the leading-order theory contains a finite cosmological constant. On the other hand, for $\Delta>-4$, the cosmological constant appears in the sub-leading theories. The remaining case of $\Delta<-4$, again leads to a trivial theory at leading-order with the first non-triviality appearing in the expansion arising in a manner analogous to the $\Delta=-4$ case. We therefore have following options of the leading order theory
\begin{eqnarray}
    \lim_{\epsilon\to0}\epsilon^{5-d}S\approx\frac{1}{16\pi G} \int_\mathcal{M} d^dx~e\begin{cases}
v^{\m\n}v^{\beta\kappa}\del_{[\kappa}h_{\n]}\del_{[\beta}h_{\m]}-2\overset{\{-4\}}{\Lambda},&\Delta=-4\\
v^{\m\n}v^{\beta\kappa}\del_{[\kappa}h_{\n]}\del_{[\beta}h_{\m]},&\Delta>-4.
    \end{cases}
\end{eqnarray}
where, $e=\sqrt{-\det(-h_{\m}h_\n+v_{\m\n})}$. We will skip further discussions of these theories here and instead refer readers to \cite{c} for a more detailed discussion. 

\medskip

Before we conclude this sub-section, let's make a few comments on the expansion and the expansion parameter $\epsilon$.\\
\newline
\underline{\textit{Comment on the expansion}:} In this subsection, we have demonstrated how we can reduce the unified decomposition of the Einstein-Hilbert action to the pre-non-relativistic decomposition of general relativity. We then further expanded the theory and extracted an action of Galilean gravity. This suggests that the expansion parameter $\epsilon$ is related to the speed of light `$c$' in the following way
\begin{eqnarray}
    \epsilon\sim c^{-1}.
\end{eqnarray}
However, one must also introduce an overall scale in the Einstein-Hilbert action to ensure correct dimension counting. For instance, in $d=4$, we can obtain this scale by noting the leading-order power in the decomposition given in \cite{c}.
\subsection{Carroll gravity}
\label{subsec:CG}
We now proceed to realize the pre-ultra-local expansion of general relativity to capture Carroll gravity from the unified decomposition of the Einstein-Hilbert action. This will be achieved by the following choice of the pair ($s,n$)
\begin{eqnarray}\label{Carroll-choice}
    s=1,\quad n=1.
\end{eqnarray}
As discussed in \cref{sec:Symmetries}, this specializes the unified local symmetry algebra to the Carroll algebra. Hence, we expect to obtain Carroll gravity at the leading order of the expansion. Further, the redefined variables in \cref{def-in-frame-fields} become
\begin{eqnarray}\label{redefined-c}
    V^{\m\n}=-V^\m V^\n,\quad V_{\m\n}=-V_{\m}V_{\n},\quad \Pi_{\m\n}=\delta_{ab}\mathcal{E}_\m^{\ a}\mathcal{E}_\n^{\ b},\quad \Pi^{\m\n}=\delta^{ab}\mathcal{E}^\m_{\ a}\mathcal{E}^\n_{\ b},
\end{eqnarray}
where we have used the notation $V^\m=-\mathcal{E}^\m_{\ t}$ and $V_\m=\mathcal{E}_\m^{\ t}$. In the limit $\epsilon\to0$ it becomes
\begin{eqnarray}
    \lim_{\epsilon\to0} V^\m=v^\m,\quad \lim_{\epsilon\to0}V_\m=v_\m.
\end{eqnarray}
We can further check that the following geometric data remains inert under the local Carroll transformation
\begin{eqnarray}
    \delta_\lambda v^\m =0,\quad \delta_\lambda h_{\m\n}=0.
\end{eqnarray}
Since $v^\m$ is the kernel for $h_{\m\n}$, i.e. $h_{\m\n}v^\n=0$, we recover the Carrollian geometry on spacetime manifold \cite{Duval:2014uoa}. Further, the orthogonality and normalization conditions read
\begin{eqnarray}
  v^\m v_\m=-1  ,\quad v_{\m}h^{\m\n}=0,\quad -v^\m v_\n+h^{\m\sigma}h_{\sigma \nu}=\delta^\m_{\ \n}.
\end{eqnarray}
We now compute the affine connection in \cref{SC-connection-Lorentz} by substituting the values of the pair $(s,n)$ (refer to Appendix~\ref{app:mapping connection}). This gives
\begin{eqnarray}\label{PUL-connection}
    \widehat{C}^\rho_{\ \m\n}=-V^\rho \del_{(\m}V_{\n)}-V^\rho V_{(\m}\pounds_V V_{\n)}+\Pi^{\rho\alpha}\Pi_{\alpha \m\n}-\Pi^{\rho\alpha}V_\n K_{\m\alpha}. 
\end{eqnarray}
To realize it in its usual form as shown in \cite{Hansen:2021fxi}, we have defined the extrinsic curvature as, $K_{\m\n}=-\frac{1}{2}\pounds_V \Pi_{\m\n}$. As a result, we reproduce the pre-ultra-local connection compatible with the Lorentzian variables $V^\mu$ and $\Pi_{\mu\nu}$. We proceed to specialize the unified decomposition of the Einstein-Hilbert action \cref{unified-EH} using \cref{Carroll-choice}. We get the following decomposition 
\begin{eqnarray}
     S\approx \frac{1}{16\pi G}\int d^dx ~E~  \Big[&&\epsilon^{-1}\left(K_{\m\n}K^{\m\n}-K^2\right)+\epsilon ~\Pi^{\m\n}\widehat R_{\m\n}\notag\\&&\quad~~~~~~~~~~~+\epsilon^{3}\left(\Pi^{\rho\beta}\Pi^{\n\alpha}\del_{[\beta}V_{\n]}\del_{[\rho}V_{\alpha]}\right)-2\epsilon^{\Delta+1} \widehat\Lambda\Big],
\end{eqnarray}
where $E=\sqrt{-\det(-V_\m V_\n+\Pi_{\m\n})}$ and $K=K_{\m\n}\Pi^{\m\n}$\footnote{It can be shown that the extrinsic curvature satisfies the following property $K_{\m\n} V^\n=0$. Therefore, we can write, $K^{\m\n}=\Pi^{\m\sigma}\Pi^{\n\beta}K_{\sigma \beta}$.}. We have successfully converted the unified decomposition to the pre-ultra-local decomposition of Einstein's gravity\footnote{The Ricci tensor term at order $\epsilon^{-1}$ drops out because one can show that $\widehat R_{\m\n}V^\nu=0$. To see this, we use the definition of the curvature tensor
\begin{eqnarray}
    [\widehat \N_\rho,\widehat \N_\n]V^\rho=\widehat R_{\ \sigma \n}V^\sigma+\widehat T^\lambda_{\rho\n}\widehat \N_\lambda V^\rho.
\end{eqnarray}
Now using the compatibility condition $\widehat \N_\sigma V^\rho=0$, this equation reduces to $R_{\ \sigma\n}V^\sigma=0$.}. Now, the cosmological constant term can be chosen as follows. First, if $\Delta=-2$, we get a finite cosmological constant in the leading-order Carroll gravity. Whereas, any $\Delta>-2$ removes it from the leading-order theory. We again omit the case $\Delta<-2$ following a similar argument to that in the previous subsections. Therefore, in the limit $\epsilon\to0$ we get
\begin{eqnarray}
    \lim_{\epsilon\to0}\epsilon S\approx\frac{1}{16\pi G} \int_\mathcal{M} d^dx~e\begin{cases}
\mathcal K_{\m\n}\mathcal{K}^{\m\n}-\mathcal{K}^2-2\overset{\{-2\}}{\Lambda},&\Delta=-2\\
\mathcal K_{\m\n}\mathcal{K}^{\m\n}-\mathcal{K}^2,&\Delta>-2,
    \end{cases}
\end{eqnarray}
where we have defined $e=\sqrt{-\det(-v_\m v_\n+h_{\m\n})}$, $\mathcal{K}_{\m\n}=-\frac{1}{2}\pounds_v h_{\m\n}$ and $\mathcal{K}=h^{\m\n}\mathcal{K}_{\m\n}$. This analysis shows how we obtain Carroll gravity from the unified expansion of Einstein's gravity by making a particular choice of the parameters $s$ and $n$. However, for a detailed discussion on this pre-ultra-local expansion of general relativity and Carroll gravity, we refer the readers to \cite{Hansen:2021fxi}.

\medskip

Let's now conclude this sub-section with the following remark and comment on the expansion and expansion parameter $\epsilon$.\\
\newline
\underline{\textit{Comment on the expansion}:} In this subsection, we have demonstrated how we can reduce the unified decomposition of Einstein-Hilbert action to the pre-ultra-local decomposition of general relativity. This implies that the expansion parameter $\epsilon$ is related to the speed of light `$c$' in the following way
\begin{eqnarray}
    \epsilon\sim c.
\end{eqnarray}
Again, we need to put an overall scale in the Einstein-Hilbert action to make the full action dimensionally consistent. For instance, in $d=4$, this scale can be recovered by identifying the leading-order power in the decomposition presented in \cite{Hansen:2021fxi}.

\section{String Carroll gravity}
\label{subsec: String Carroll gravity}
In this section, we discuss the specialization of the unified decomposition and expansion for a string Carroll geometry. This geometry has been proposed to appear in the near-horizon of generic non-extremal black holes (BHs)\cite{Bagchi:2023cfp,Bagchi:2024rje}(see also \cite{Fontanella:2022gyt} for a related discussion of non-Lorentzian structures in near-horizon geometries). Although we do not yet have a rigorous proof of this identification, it has consistently appeared in many known examples; see \cite{Bagchi:2026qpi}. Therefore, the string Carroll gravity equations one obtains here must capture the dynamics and constraints satisfied by the near-horizon geometry of non-extremal BHs. As explicit evidence, we consider an example of a non-extremal BH in \cref{Kerr-BH} and a black brane in \cref{BTZ-BH}, and show that their near-horizon geometries satisfy the string Carroll gravity equations. This supports the claim made in \cite{Bagchi:2023cfp,Bagchi:2024rje,Banerjee_2026,Bagchi:2026qpi}. Subsequently, our construction provides a systematic way to examine and capture the departures in the dynamics of non-extremal BHs from the strict near-horizon limit. Therefore, in this case, the scaling parameter $\epsilon$ is naturally related to the control parameter of nearness from the horizon. A detailed analysis of this departure is deferred to future work.

\medskip

The local string Carroll algebra appears from the unified algebra for the following choice of the parameter 
\begin{eqnarray}
    s=1,\quad n=2.
\end{eqnarray}
This implies the following form of the redefined variables \cref{def-in-frame-fields}
\begin{subequations}
    \begin{align}
        & V^{\m\n}=-\mathcal{E}_{\ 0}^\m \mathcal{E}_{\ 0}^\n+\mathcal{E}_{\ 1}^\m \mathcal{E}_{\ 1}^\n,\quad V_{\m\n}=-\mathcal{E}^{\ 0}_\m \mathcal{E}^{\ 0}_\n+\mathcal{E}^{\ 1}_\m \mathcal{E}^{\ 1}_\n,\\
        &\Pi_{\m\n}=\mathcal{E}_\m^{\ a}\mathcal{E}_\n^{\ b}\delta_{ab},\quad \Pi^{\m\n}=\mathcal{E}^\m_{\ a}\mathcal{E}^\n_{\ b}\delta^{ab}.
    \end{align}
\end{subequations}
Here, the lowercase Latin indices run over $d-2$ spatial directions. As a result, $\Pi_{\mu\nu}$ carries two vanishing eigenvalues.

\medskip

In the Carroll and Galilean gravity cases, we expressed all relevant geometric objects and the action in terms of the vielbeins $(V^\mu, V_\mu)$ and $(\Pi_{\mu}, \Pi^\mu)$, respectively. An analogous construction can also be done here using $(\mathcal{E}^{\ 0}_\mu, \mathcal{E}^{\ 1}_\mu)$ and $(\mathcal{E}_{\ 0}^\mu, \mathcal{E}_{\ 1}^\mu)$. However, we will avoid working explicitly with these quantities, as the convention adopted here will be more convenient for discussing explicit examples later in \cref{Kerr-BH} and \cref{BTZ-BH}. Therefore, for this case, we will retain the form of the unified connection \cref{SC-connection-Lorentz} and the decomposition of the Einstein-Hilbert action \cref{unified-EH} stays the same
\begin{align}
     S\approx \frac{1}{16\pi G}\int d^dx ~E~ \Big[&\epsilon^{-2}\left(\Pi^{\rho\alpha}V^{\beta\kappa}V^{\m\n}\Pi_{\kappa\rho\n}\Pi_{\m\alpha\beta}\right)+\left(2V^{\sigma \lambda}\Pi_{\sigma \beta\m}\Pi_{\lambda \rho \n}\Pi^{\m[\beta}\Pi^{\nu]\rho}+V^{\m\n}\widehat R_{\m\n}\right)\notag\\
    &+\epsilon^{2}\Pi^{\m\n}\widehat R_{\m\n}+\epsilon^{4}\left(\Pi^{\rho\beta}V_{\beta\m\n}V^{\m \alpha}\Pi^{\n \kappa}V_{\kappa \rho \alpha}\right)-2\epsilon^{\Delta+2} \widehat\Lambda\Big]
\end{align}
We now make the choice of $\Delta$ to keep or scale out the contribution from the cosmological constant in the leading-order theory. As a result, we obtain the following theories at the leading order (skipping the trivial case again)
\begin{eqnarray}\label{sc-gravity-action}    &&\notag\lim_{\epsilon\to 0}\epsilon^2 S\approx \widetilde S=\frac{1}{16\pi G}\int_\mathcal{M} d^dx~e~\mathcal{L}^{(0)} \,,\\ &&\text{where},\quad \mathcal{L}^{(0)}=\begin{cases}
h^{\rho \alpha
} v^{\beta\kappa}v^{\m\n}h_{\kappa\rho \nu}h_{\m\alpha \beta}-2\overset{\{-4\}}{\Lambda},&\Delta=-4\\
h^{\rho \alpha
} v^{\beta\kappa}v^{\m\n}h_{\kappa\rho \nu}h_{\m\alpha \beta},&\Delta>-4.
    \end{cases}
\end{eqnarray}
Now, we can proceed to compute the equation of motion of the leading-order theory. It essentially encodes the equations for constraints and dynamics for the string Carroll geometry. We will vary the action \cref{sc-gravity-action} with respect to $v^{\m\n}$ and $h^{\m\n}$. In order to compute this variation, we can use the following relations of variational calculi
\begin{subequations}
    \begin{align}
        \delta v_{\mu\nu} &= - v_{\mu\rho}v_{\nu\lambda} \delta v^{\rho\lambda} - 2 h_{\rho(\mu} v_{\nu)\lambda} \delta h^{\rho\lambda} \,, \\
        \delta h_{\mu\nu} &= - 2 h_{\rho(\mu} v_{\nu)\lambda} \delta v^{\rho\lambda} - h_{\mu\rho}h_{\nu\lambda} \delta h^{\rho\lambda} \,.
    \end{align}
\end{subequations}
As a result, we get the following variation of \cref{sc-gravity-action}\footnote{We compute it for the $\Delta=-4 $ case because the result can be extended to $\Delta>-4$ by simply setting $\overset{\{-4\}}{\Lambda}=0$.}
\begin{eqnarray}\label{SC-gravity-variation}
    \delta \widetilde S\approx \frac{1}{16\pi G}\int_\mc M \left[\mc E^{(v)}_{\mu\nu}\delta v^{\mu\nu}+\mathcal{E}^{(h)}_{\mu\nu}\delta h^{\m\n}\right] e~d^{d}x,
\end{eqnarray}
where the equations of motion are given by, $\mc E^{(v)}_{\mu\nu}=0$ and $\mathcal{E}^{(h)}_{\mu\nu}=0$. They are given by
\begin{subequations}\label{SC-gravity-eom 0}
    \begin{align}
        &\label{eom-1}\mc E^{(v)}_{\mu\nu}=- \frac{1}{2}\mathcal{L}^{(0)}v_{\m\n}-2h^{\rho \alpha
} v^{\beta \kappa}\del_{[\n}h_{\kappa]\rho}\del_{[\m}h_{\beta] \alpha}+\frac{2}{e} h_{\alpha(\m }v_{\n)\beta}\del_\sigma(e h^{\rho \alpha }v^{\beta\kappa}v^{\sigma \lambda}\del_{[\kappa}h_{\lambda]\rho}),\\
        &\mathcal{E}^{(h)}_{\mu\nu}=- \frac{1}{2}\mathcal{L}^{(0)}h_{\m\n}+v^{\beta\kappa}v^{\gamma\sigma}h_{\kappa\m \sigma}h_{\gamma\nu \beta}+2 v^{\sigma\lambda}\del_\sigma v^{\beta\kappa}\del_{[\kappa}h_{\lambda](\m}h_{\n)\beta}.
    \end{align}
\end{subequations}
Before proceeding further, we emphasize that the equations derived above do not depend on the specific values of $s$ and $n$, provided the action is finite. Hence, they remain valid for all other admissible choices of $s$ and $n$, excluding the Carroll, Lorentzian, and Euclidean cases. As a quick consistency check, \cref{SC-gravity-eom 0} can be shown to reproduce the Galilean gravity equations computed in \cite{c}. This is done by substituting the limiting version of \cref{galilean-vielbeins} into \cref{SC-gravity-variation}. To see this explicitly, refer to Appendix \ref{appen:galilean gravity}. Now, using the trace of \cref{eom-1} with $v^{\m\n}$, we can rewrite \cref{SC-gravity-eom 0} as
\begin{subequations}\label{SC-gravity-eom}
    \begin{align}
     &h^{\rho \alpha
} v^{\beta\kappa}v^{\m\n}h_{\kappa\rho \nu}h_{\m\alpha \beta}=-2\overset{\{-4\}}{\Lambda},\label{eq-1-sc}\\
&h^{\rho \alpha
} v^{\beta \kappa}\del_{[\kappa}h_{\nu]\rho}\del_{[\beta}h_{\m] \alpha}+\frac{1}{e} h_{\alpha(\m }v_{\n)\beta}\del_\sigma[e h^{\rho \alpha }v^{\beta\kappa}v^{\sigma \lambda}\del_{[\lambda}h_{\kappa]\rho}]=\overset{\{-4\}}{\Lambda}v_{\m\n},\label{eq-2-sc}\\
&v^{\sigma\lambda}\del_\sigma v^{\beta\kappa}\del_{[\lambda}h_{\kappa](\m}h_{\n)\beta}-\frac{1}{2}v^{\beta\gamma}v^{\kappa\sigma}h_{\kappa\m \beta}h_{\gamma\nu \sigma}=\overset{\{-4\}}{\Lambda}h_{\m\n},\label{eq-3-sc}
    \end{align}
\end{subequations}
for string Carroll gravity. This set of equations captures the dynamics and constraints that a string Carroll geometry must satisfy, obtained from the $\epsilon\to0$ limit of a geometry satisfying Einstein's field equation.
\medskip

\underline{\textit{Comment on the expansion }:} As we discussed before, this expansion is going to map the near-horizon expansion of a non-extremal BH. The smallness parameter $\epsilon$ in this case will be related to the nearness parameter $\varepsilon$ to the horizon. We will now see some examples of the near-horizon geometries and check their consistency with the string Carroll gravity.
\subsection{ Near-horizon of Plebanski-Demianski BHs ($Q_{e,m}=0$)}\label{Kerr-BH}
In this part of the section, we will examine the near-horizon geometry of a seven-parameter family of stationary and axisymmetric black hole solutions, known as the Plebanski-Demianski family \cite{Plebanski:1976gy}, which contains Schwarzschild, Kerr, and Reissner-Nordström as special cases. Showing that its near-horizon geometry satisfies the string Carroll gravity equations \cref{SC-gravity-eom}, will qualify all these geometries as a consistent solution. However, we must set the electric charge $(Q_e)$ and magnetic charge $(Q_m)$ to zero, since this work explores only vacuum Einstein's gravity with a finite cosmological constant. Hence, we do not expect BHs with finite $Q_e$ and $Q_m$ to satisfy \cref{SC-gravity-eom}.
\medskip

The remaining five parameters are mass ($M$), angular momentum ($J=aM$), cosmological constant ($\Lambda$), Taub-NUT charge ($l$) and acceleration ($\alpha$). The metric of the Plebanski-Demianski family in the Boyer-Lindquist coordinates is given as 
\begin{align}
     g_{\m\n}dx^\m dx^\n&=\frac{1}{\Omega(r, \theta)^2}\Bigg(-\frac{\widetilde\Delta(r)}{\Sigma(r, \theta)}\Big(dt-\Gamma(\theta)~d\phi\Big)^2+\frac{\Sigma(r, \theta)}{\widetilde\Delta(r)}dr^2+\frac{\Sigma(r, \theta)}{P(\theta)}d\theta^2   \nonumber\\
    &\quad +\frac{P(\theta)\sin^2\theta}{\Sigma(r, \theta)}\Big(a~dt-(r^2+a^2+l^2)~d\phi\Big)^2\Bigg)\,. \label{eq:PD Metric}
\end{align}
The metric functions and constants are given as 
\begin{subequations}
    \begin{align}
    P(\theta) =&~1+ak_1\cos\theta+a^2k_2\cos^2\theta\,,\\
    \Omega(r,\theta) =&~1-\alpha(l+a\cos\theta)~r\,,\\
    \Sigma(r, \theta) =&~r^2+(l+a\cos\theta)^2\,,\\
    \widetilde\Delta(r) =&~ \left(k\left(2 \alpha  l r+1\right)-2 M r+\frac{kr^2}{a^2-l^2}\right)\Big(1-\alpha(a+l)r\Big) \Big(1+\alpha(a-l)r\Big)\nonumber\\
    &~-\frac{\Lambda}{3}r^2 \Big(2 \alpha  l r \left(a^2-l^2\right)+a^2+3 l^2+r^2\Big)\,,\\
    \Gamma(\theta) =&~a\sin^2\theta-2l\cos\theta\,,\\
    k=&~(a^2-l^2)\frac{1+2M\alpha l-\Lambda l^2}{1+3\alpha^2l^2(a^2-l^2)}\,,\\
    k_1=&~4\alpha^2lk-2\alpha M+4l\frac{\Lambda}{3}\,,\\
    k_2=&~\alpha^2 k+\frac{\Lambda}{3}\,.
    \end{align}
\end{subequations}
The horizons of the black hole are the roots of the metric function $\widetilde\Delta$. If the parameters are set appropriately, the outermost horizon would be the event horizon. We represent its radial coordinate location with $r_h$, hence, $\widetilde\Delta (r_h)=0$. This becomes a crucial input for the expansion of geometry, as it specifies that it is performed in the vicinity of the horizon.

\medskip
To probe the near-horizon region of the line element \cref{eq:PD Metric}, we introduce a small dimensionless parameter $\varepsilon$\footnote{The nearness parameter $\varepsilon$ identifies as
\begin{eqnarray}
    \varepsilon=\epsilon^2.
\end{eqnarray} } via following coordinate transformation
\begin{equation}
    r=r_h+\varepsilon\rho^2 \,,
\end{equation}

where $\rho$ is treated as a secular coordinate. Now the metric \cref{eq:PD Metric} admits the following string Carroll decomposition \footnote{Subsequently, the $\theta$ dependence of $\Omega,\, \Sigma,\, P$ is not explicitly mentioned for brevity.} 
\begin{subequations}\label{eq:PD metric split}
\begin{align}
    g=~&\varepsilon V+\Pi\,,\\
    \Pi=~&\frac{1}{\Omega^2}\Bigg(\frac{\Sigma}{P}d\theta^2+\frac{P\sin^2\theta}{\Sigma}\Big(a~dt-((r_h + \varepsilon  \rho^2)^2 +a^2+l^2)~d\phi\Big)^2\Bigg)\,,\\
    V=~&\frac{1}{\Omega^2}\Bigg(-\frac{\left( \widetilde\Delta / \varepsilon \right) }{\Sigma}\Big(dt-\Gamma~d\phi\Big)^2+4 \frac{\Sigma \rho^2}{\left( \widetilde\Delta / \varepsilon \right)}d\rho^2\Bigg)\,.
\end{align}    
\end{subequations}
This decomposition is done from making following observation $\Omega=\mathcal{O}(1)$, $\Sigma=\mathcal{O}(1)$ and $\widetilde\Delta=\mathcal{O}(\varepsilon)$.  Similarly, the inverse metric $g^{-1}$ can be split into $g^{-1}=\frac{1}{\varepsilon
}V^{-1}+\Pi^{-1}$, where
\begin{subequations}
    \begin{align}
        V^{-1}=~&\Omega^2\Bigg(-\frac{\Sigma\Big((a^2+l^2+(r_h + \varepsilon  \rho^2)^2  )\del _t+a~\del_ \phi\Big)^2}{\left(a^2+l^2+(r_h + \varepsilon  \rho^2)^2 -a\Gamma\right)^2 \left(  \widetilde\Delta / \varepsilon \right)}+\frac{\left(\widetilde\Delta / \varepsilon \right)}{ 4 \rho^2 \Sigma}\del_\rho^2\Bigg)\,,\\
        \Pi^{-1}=~&\Omega^2\Bigg(\frac{P}{\Sigma}\del_\theta^2+\frac{\Sigma\csc^2\theta}{\left(a^2+l^2+(r_h + \varepsilon  \rho^2)^2 -a\Gamma\right)^2P}\Big(\Gamma~\del_t + \del_\phi\Big)^2\Bigg)\,.
    \end{align}
\end{subequations}
Now, using this decomposition, we can extract the data for the string Carroll geometry and its inverse\footnote{\label{ft: plebanski-demianski local invariance}Although the inverse data is not invariant under the local transformations \cref{local-unified-transformation}, we may nevertheless choose particular data consistent with \cref {LO-ortho-norm} and verify whether it satisfies the equations of string Carroll gravity. Since, by construction, these equations are invariant under local transformations, any other local choice of the inverse data will also satisfy them.}
\begin{subequations}\label{near-kerr-1}
\begin{align}
h_{\mu\nu}dx^\mu dx^\nu=~&\frac{1}{\Omega(r_h)^2}\Bigg(\frac{\Sigma(r_h)}{P}d\theta^2+\frac{P\sin^2\theta}{\Sigma(r_h)}\Big(a~dt-(r_h^2+a^2+l^2)~d\phi\Big)^2\Bigg)\,,\\
v_{\mu\nu}dx^\mu dx^\nu=~&\frac{1}{\Omega(r_h)^2}\Bigg(-\frac{\rho ^2 \partial_r \widetilde\Delta\Big|_{r=r_h}}{\Sigma \left(r_h\right)}\Big(dt-\Gamma d\phi\Big)^2+  \frac{4 \Sigma \left(r_h\right)}{\partial_r \widetilde\Delta\Big|_{r=r_h} } d\rho^2\Bigg)\,,\\
h^{\m\n}\del_\m\del_\n=~& \Omega (r_h )^2 \left(\frac{P  }{\Sigma (r_h )}\partial_\theta^2 + \frac{\csc ^2(\theta ) \Sigma \left(r_h\right) }{P \left(a^2 - a \Gamma + r_h^2+l^2\right){}^2} (\partial_\phi + \Gamma  \partial_t)^2 \right) ,\\
v^{\m\n}\del_\m\del_\n=~& -\Omega(r_h)^2\Bigg(\frac{\Sigma \left(r_h\right) }{ \rho ^2 \partial_r \widetilde\Delta\Big|_{r=r_h} \left(a^2-a \Gamma +r_h^2+l^2\right){}^2}\left( \left(a^2+l^2+r_h^2\right)\partial_t+a \partial_\phi\right)^2 \notag\\&\quad ~~~~~~~~~~~~~~~~~~~~~~~~~~~~~-  \frac{\partial_r \widetilde\Delta\Big|_{r=r_h} }{4 \Sigma \left(r_h\right)}\partial_\rho ^2 \Bigg).
\end{align}
\end{subequations}

On substituting \cref{near-kerr-1} in \cref{SC-gravity-eom}, one can see that this data indeed satisfies the dynamics and constraints of string Carroll gravity with $\overset{\{-4\}}{\Lambda}=0$. This is precisely what one should expect since the string Carroll geometry in \cref{near-kerr-1} is obtained without any scaling of $\Lambda$. As a result, it corresponds to the case $\Delta=0$ of the string Carroll expansion of gravity. Hence, the near-horizon geometry of Plebanski-Demianski BH with $Q_e=Q_m=0$ is a dynamical solution of string Carroll gravity.

\subsection{ Near-horizon of a $4D$ black brane}\label{BTZ-BH}
We now compute the near-horizon geometry of a black brane metric. A 4D black brane metric is given by
\begin{eqnarray}\label{bb-metric}
    g_{\m\n}dx^\m dx^\n=-\frac{r^2}{l^2}\left(1-\frac{r_h^3}{r^3}\right) dt^2+\frac{l^2}{r^2}\left(1-\frac{r_h^3}{r^3}\right)^{-1}dr^2+\frac{r^2}{l^2}\left(dx^2+dy^2\right),
\end{eqnarray}
where the AdS radius is related to the cosmological constant as $l^{-2}=-\Lambda/3$. To compute the near-horizon geometry of this metric, we make the following coordinate transformation
\begin{eqnarray}
    r=r_h+\varepsilon \rho \Rightarrow \rho=\frac{r-r_h}{\varepsilon}.
\end{eqnarray}
Substituting this coordinate transformation in \cref{bb-metric} and then sending $\varepsilon \to0$, we get the following data for string Carroll geometry
\begin{subequations}
\begin{align}
        &\lim_{\varepsilon\to0}g=h_{\m\n}dx^\m dx^\n=\frac{r_h^2}{l^2}(dx^2+dy^2),\\
        &\lim_{\varepsilon\to 0}\varepsilon g^{-1}=v^{\m\n}\del_\m\del_\n=-\frac{l^2}{3 r_h \rho}\del_t^2+\frac{3 r_h \rho}{l^2}\del_\rho^2.        
    \end{align}
\end{subequations}
We again borrow the inverse data from the expansion (refer to footnote~\eqref{ft: plebanski-demianski local invariance} for the rationale of this choice)
\begin{subequations}\label{near-kerr-2}
    \begin{align}
        &h^{\m\n}\del_\m \del_\n=\frac{l^2}{r_h^2 }( \del_x^2+\del_y^2),\\
        &v_{\m\n}dx^\m dx^\n=-\frac{3 r_h \rho}{l^2}dt^2+\frac{l^2}{3 r_h \rho}d\rho^2.
    \end{align}
\end{subequations}
It can be verified that this data is consistent with \cref{LO-ortho-norm} and satisfies the string Carroll gravity equations \cref{SC-gravity-eom} with $\overset{\{-4\}}{\Lambda}=0$, again in this case $\Delta=0$. \\
\newline
With this, we establish that both the near-horizon non-extremal Plebanski-Demianski geometry (with $Q_e=Q_m=0$) and the black brane in 4D are consistent solutions of string Carroll gravity without a cosmological constant. The same conclusion for black branes extends straightforwardly to their higher-dimensional analogs.
\section{Discussion and outlook}
\label{sec:Discussion}
In this paper, we presented a unified framework for expanding Einstein's theory of gravity. Our construction exploits a unified notion of flat geometry together with its associated symmetry algebra in the tangent space. The central idea is to formulate a covariant structure that remains insensitive to certain details of the underlying construction. This framework is characterized by a two-parameter family of data. We represented them by $(s,n)$ throughout. A specific choice of $(s,n)$ selects a particular flat metric and symmetry algebra in the tangent space. We showed that the resulting geometry and their respective symmetry algebra can be broadly classified into four classes: Galilean, Carrollian, Lorentzian, and Euclidean. For example, in $d=4$, the possible choices of ($s,n$) are classified as follows\footnote{In our convention, we call an algebra containing $p$ copies of the Galilean or Carroll sub-algebras (along with rotations and Lorentz boosts) a $p-\text{Galilei}$ or $p-\text{Carroll}$ algebra, respectively. }
\begin{table}[h]
\centering
\renewcommand{\arraystretch}{1.5}
\begin{tabular}{|>{\centering\arraybackslash}m{1.2cm}
                |>{\centering\arraybackslash}m{2.2cm}
                |>{\centering\arraybackslash}m{2.2cm}
                |>{\centering\arraybackslash}m{2.2cm}
                |>{\centering\arraybackslash}m{2.2cm}
                |>{\centering\arraybackslash}m{2.2cm}|}
\hline
\diagbox[
  width=\dimexpr1.2cm+2\tabcolsep\relax,
  height=1cm,
  innerleftsep=4pt,
  innerrightsep=4pt
]{$s$}{$n$}
&
\textbf{0} &
\textbf{1} &
\textbf{2} &
\textbf{3} &
\textbf{4}
\\
\hline
\textbf{0} & Lorentzian & 3-Galilei & String Galilei & Galilean & Euclidean\\
\hline
\textbf{1} & Euclidean & Carroll & String Carroll & 3-Carroll & Lorentzian\\
\hline
\end{tabular}
\caption{Classification of local symmetry algebra obtained by fixing the parameters $(s,n)$ in $d=4$ unified flat geometry.}
\label{tab:geometry-classification}
\end{table}

\medskip

Using this covariant setup in terms of an arbitrary $(s,n)$-pair in the tangent space, we expanded Einstein's theory of gravity. Subsequently, we showed how the resulting expansion encompasses Galilean, Carrollian, and string Carroll gravity as particular realizations. We also emphasized that the physical interpretation of both the expansion and the expansion parameter is fixed by the choice of $(s,n)$. For instance, in the case of Galilean and Carroll gravity, the expansion parameter is naturally identified with the speed of light `$c$'. Whereas, in string Carroll gravity, it maps to the nearness parameter of a non-extremal black hole to its event horizon `$\varepsilon$'. 

\medskip

The possibility of such a unified framework naturally leads us to ask the following question: Can we learn common lessons that transcend all realizations of the unified structure? Since Galilean, Carrollian, Lorentzian, and Euclidean geometries emerge as different choices of the same underlying framework. Therefore, it is natural to expect that certain geometric, algebraic, or dynamical features may admit some common notions. Understanding and classifying which features are universal and which are specific to a particular case could provide deeper insight into the various relations among these seemingly distinct theories. Such an investigation may also reveal new kinds of dualities; for instance, in $d=2$, Galilean ($s=0$ and $n=1$) and Carroll ($s=1$ and $n=1$) theories are dual to each other by a flip of the notion between time and space.

\medskip

In this paper, we have restricted our attention to Einstein's gravity. A particularly promising avenue for future investigation is the inclusion of higher-derivative corrections, which are known to arise in an effective description of quantum gravity and string theory (refer to \cite{cardona2025higherordernewtoncartangravity,Lescano:2025xzs,Lescano:2026hxd} for a non-relativistic version of such a theory). It would be interesting to determine whether the unified framework developed in this work admits a natural extension to theories of gravity with higher-derivative corrections. In such theories, the correction terms are typically weighted by a dimensionless parameter $\alpha'$

\begin{equation}
S=
\frac{1}{16 \pi G}
\int d^d x\,\sqrt{-g}
\left[
R
+\alpha'
\left(
a_1 R_{\mu\nu\rho\sigma}R^{\mu\nu\rho\sigma}
+a_2 R_{\mu\nu}R^{\mu\nu}
+a_3 R^2
\right)
+\mathcal{O}\!\left((\alpha')^2\right)
\right].
\end{equation}

This raises the intriguing possibility of an interplay between the two expansion parameters, namely $\alpha'$ and the unified expansion parameter $\epsilon$. Depending on how these parameters are scaled relative to one another, one can reach different sectors of the theory in the limit, although one must still fix $s$ and $n$ in the end to specialize the theory. 

\medskip

In this work, we have focused exclusively on a theory of gravity. A natural extension of the present framework would be to explore whether a similar unified description can be developed for other physical theories, such as field theory, hydrodynamics, and string theory. Given the broad range of contexts in which non-Lorentzian structures arise, it is a compelling direction to explore. In conclusion, it is remarkable that a single unified framework can capture a variety of geometries and symmetry algebras. We hope that the setup developed in this work will provide a useful starting point for uncovering deeper connections among various classes of theories.
\acknowledgments
We thank Arjun Bagchi for his valuable insights in the project, for carefully reading the draft, and for providing detailed feedback. We thank Jos\'e M. Mart\'in-Garc\'ia for the well-documented \texttt{xTensor} package \cite{xTensor}, part of the \texttt{xAct} suite, which we used in some of our tensor computations. Arkachur would like to thank Sharang R. Iyer and Adil Imam for their useful comments and discussions. PS would also like to thank Kedar S. Kolekar for introducing him to this topic. PS is grateful to the organizers of the Solvay Workshop on ``\textit{Carrollian Physics and Geometry}'', held at ULB Campus Plaine, Brussels, for the opportunity to participate. The insightful talks and discussions during the workshop provided valuable perspectives that benefited this project. Arkachur and PS are supported by the IIT Kanpur Institute Assistantship.

\appendix
\section{Derivation of unified compatible connection}
\label{app:derviation-of-connection}
In this appendix, we compute an affine connection satisfying \cref{SC-connection-Lorentz}. We proceed analogously to Appendix (B.2) in ref.~\cite{Hansen:2021fxi}, where the derivation is done in terms of the frame fields
\begin{eqnarray}
    \mathcal{E}^{\ A}_{\m}dx^\m=\mathcal{E}^A,\quad \mathcal{E}^{\ a}_{\m}dx^\m=\mathcal{E}^a,\quad \mathcal{E}_{\ A}^{\m}\del_\m=\mathcal{E}_A,\quad \mathcal{E}_{\ a}^{\m}\del_\m=\mathcal{E}_a.
\end{eqnarray}
We first compute the constraints on spin-connections $\Omega^A_{\ B},~\Omega^A_{\ b},~\Omega^a_{\ B}$ and $\Omega^a_{\ b}$ imposed by the compatibility conditions \cref{SC-compatibility-Lorentzian}. We then turn off all torsion components that do not constrain the geometry and solve for the spin connections in terms of the vielbeins. Finally, we simply invert the vielbein postulates
\begin{subequations}\label{vielbein-postulates}
    \begin{align}
        \label{postulate-1}0&=\del_\m \mathcal{E}_\n^{\ A}-\widehat{C}^\rho_{\ \m\n}\mathcal{E}_\rho ^{\ A}+\Omega^{\ A}_{\m\  B}\mathcal{E}^{\ B}_\n+\Omega^{\ A}_{\m\ b}\mathcal{E}^{\ b}_\n,\\  0&=\del_\m \mathcal{E}_\n^{\ a}-\widehat{C}^\rho_{\ \m\n}\mathcal{E}_\rho ^{\ a}+\Omega^{\ a}_{\m\ B}\mathcal{E}^{\ B}_\n+\Omega^{\ a}_{\m\  b}\mathcal{E}^{\ b}_\n,\label{postulate-2}\\
        0&=\del_\m \mathcal{E}^\n_{\ A}+\widehat{C}^\n_{\ \m\rho}\mathcal{E}^\rho _{\ A}-\Omega^{ \ B}_{\m \  A}\mathcal{E}_{\ B}^\n-\Omega^{   \ b}_{\m \  A}\mathcal{E}_{\ b}^\n,\label{postulate-3} \\ 0&= \del_\m \mathcal{E}^\n_{\ a}+\widehat{C}^\n_{\ \m\rho}\mathcal{E}^\rho _{\ a}-\Omega^{ \ B}_{\m\  a}\mathcal{E}_{\ B}^\n-\Omega^{\   b}_{\m \   a}\mathcal{E}_{\ b}^\n,\label{postulate-4}
    \end{align}
\end{subequations}
to write the affine connection in terms of vielbeins.

\medskip

Let's start by expressing the compatibility conditions in terms of the frame fields using the definitions in \cref{def-in-frame-fields}
\begin{eqnarray}
    \widehat\N_\sigma V^{\m\n}=\widehat \N_\sigma \left(\eta^{AB}\mathcal{ E}^\m_{\ A}\mathcal{ E}^\n_{\ B}\right)=0.
\end{eqnarray}
Using the vielbein postulates \cref{postulate-3}, this can be written as
\begin{eqnarray}\label{frame-compatibility-1}
    \eta^{AB}\mathcal{ E}^\n_{\ B}\left(\Omega_{\sigma \  A}^{\   C}\mathcal{ E}^\m_{\ C}+\Omega_{\sigma \  A}^{\  c}\mathcal{ E}^\m_{\ c}\right)+\eta^{AB}\mathcal{ E}^\m_{\ A}\left(\Omega_{\sigma \  B}^{\   C}\mathcal{ E}^\n_{\ C}+\Omega_{\sigma  \ B}^{\   c}\mathcal{ E}^\n_{\ c}\right)=0.
\end{eqnarray}
Similarly, the second compatibility condition reads
\begin{eqnarray}
    \widehat \N_\sigma \Pi_{\m\n}=\widehat \N_\sigma\left(\eta_{ab}\mathcal{E}^{\ a}_\m\mathcal{E}^{\ b}_\n\right)=0,
\end{eqnarray}
and using \cref{postulate-2} can be rewritten as
\begin{eqnarray}\label{frame-compatibility-2}
   \eta_{ab}\mathcal{E}_\n^{\ b}\left(\Omega_{\s\  C}^{\ a}\mathcal{E}_\m^{\ C}+\Omega_{\s\  c}^{\ a}\mathcal{E}_\m^{\ c}\right)+\eta_{ab}\mathcal{E}_{\m}^{\ a}\left(\Omega_{\s\  C}^{\ b}\mathcal{E}_\n^{\ C}+\Omega_{\s\  c}^{\ b}\mathcal{E}_\n^{\ c}\right)=0.
\end{eqnarray}
Now take the projection of \cref{frame-compatibility-1} and \eqref{frame-compatibility-2} along various combinations of frame bases and compute the conditions satisfied by the spin-connections. All non-trivial projections are
\begin{subequations}
    \begin{align}
        &\mathcal{E}_\m^{\ A} \mathcal{E}_\n^{\ B}\widehat \N_\sigma V^{\m\n}=0,\quad  \mathcal{E}_\m^{\ A} \mathcal{E}_\n^{\ b}\widehat \N_\sigma V^{\m\n}=0,\\
        & \mathcal{E}^\m_{\ A} \mathcal{E}^\n_{\ b}\widehat \N_\sigma \Pi_{\m\n}=0,\quad  \mathcal{E}^\m_{\ a} \mathcal{E}^\n_{\ b}\widehat \N_\sigma \Pi_{\m\n}=0.
    \end{align}
\end{subequations}
Solving these equations, we obtain the following conditions
\begin{eqnarray}
    \Omega^{AB}=-\Omega^{BA},\quad \Omega^{ab}=-\Omega^{ba},\quad \Omega^a_{\ A}=0.
\end{eqnarray}
Now we can use these conditions as an input to determine the components of the torsion two-forms: $T^A=(d\mathcal{E})^A+\Omega^{ A}_{\ B }\wedge\mathcal{E}^{ B}+\Omega^{ A}_{ \ b }\wedge\mathcal{E}^{ b}$ and $T^a=(d\mathcal{E})^a+\Omega^{ a}_{ \ B }\wedge\mathcal{E}^{B}+\Omega^{ a}_{\ b }\wedge\mathcal{E}^{b}$ \footnote{We are using the following notation for the exterior derivative $(d\mathcal{E})_{\m\n}=\del_\m \mathcal{E}_\n-\del_\n \mathcal{E}_\m$.}. We compute the various projections of the torsion tensor, given by
\begin{subequations}\label{torsion-2-form}
    \begin{align}
       &T^A_{\ \m\n} \mathcal{E}^{ \m}_{\ C}\mathcal{E}^{ \n}_{\ B}=(d\mathcal{ E})^A_{\ \m\n}\mathcal{E}^{ \m}_{\ C}\mathcal{E}^{ \n}_{\ B}+\Omega^{\ A}_{\m\ B}\mathcal{E}^{ \m}_{\ C}-\Omega^{\ A}_{\m\ C}\mathcal{E}^{ \m}_{\ B},\\
       &\label{projection-spin-connection}T^A_{\ \m\n} \mathcal{E}^{ \m}_{\ C}\mathcal{E}^{ \n}_{\ b}=(d\mathcal{ E})^A_{\ \m\n}\mathcal{E}^{ \m}_{\ C}\mathcal{E}^{ \n}_{\ b}-\Omega^{\ A}_{\m \ C}\mathcal{E}^{ \m}_{\ b}+\Omega_{\m\ b}^{\ A}\mathcal{E}^\m_{\ C},\\
       &T^A_{\ \m\n} \mathcal{E}^{ \m}_{\ c}\mathcal{E}^{ \n}_{\ b}=(d\mathcal{ E})^A_{\ \m\n}\mathcal{E}^{ \m}_{\ c}\mathcal{E}^{ \n}_{\ b}+\Omega^{\ A}_{\m \ b}\mathcal{E}^\m_{\ c}-\Omega^{\ A}_{\m \ c}\mathcal{E}^\m_{\ b},\\
       &T^a_{\ \m\n}\mathcal{E}^{ \m}_{\ C}\mathcal{E}^{ \n}_{\ B}=(d\mathcal{E})^a_{\ \m\n}\mathcal{E}^{ \m}_{\ C}\mathcal{E}^{ \n}_{\ B},\\
       &T^a_{\ \m\n}\mathcal{E}^{ \m}_{\ C}\mathcal{E}^{ \n}_{\ b}=(d\mathcal{E})^a_{\ \m\n}\mathcal{E}^{ \m}_{\ C}\mathcal{E}^{ \n}_{\ b}+\Omega^{\ a}_{\m \ b}\mathcal{E}_{\ C}^{ \m},\\
       &T^a_{\ \m\n}\mathcal{E}^{ \m}_{\ c}\mathcal{E}^{ \n}_{\ b}=(d\mathcal{E})^a_{\ \m\n}\mathcal{E}^{ \m}_{\ c}\mathcal{E}^{ \n}_{\ b}+\Omega^{\ a}_{\m \ b}\mathcal{E}_{\ c}^{ \m}-\Omega^{\ a}_{\m\  c}\mathcal{E}_{\ b}^{ \m}.
    \end{align}
\end{subequations}
From these components, it is straightforward to see that setting the following components to zero imposes nontrivial constraints on geometry\footnote{Note that this statement holds for the unified geometry, although it does not hold true once one specializes to the Lorentzian or Euclidean class.}
\begin{subequations}\label{Torsion-zero-constraints}
    \begin{align}
         &T^a_{\ \m\n}\mathcal{E}^{ \m}_{\ C}\mathcal{E}^{ \n}_{\ B}=(d\mathcal{E})^a_{\ \m\n}\mathcal{E}^{ \m}_{\ C}\mathcal{E}^{ \n}_{\ B},\\& T_{a\m\n}\mathcal{E}^\n_{\ b}\mathcal{E}^\m_{\ C}+T_{b\m\n}\mathcal{E}^\n_{\ a}\mathcal{E}^\m_{\ C}=(d\mathcal{E})_{a\m\n}\mathcal{E}^\n_{\ b}\mathcal{E}^\m_{\ C}+(d\mathcal{E})_{b\m\n}\mathcal{E}^\n_{\ a}\mathcal{E}^\m_{\ C}.
    \end{align}
\end{subequations}
This shows in general that the unified geometry (obtained from $\epsilon\to0$ of this geometry) carries an \textit{intrinsic} torsion for a general compatible connection. Therefore, instead of setting these components to zero, which would impose constraints on the geometry, we leave them unfixed and set the remaining components to zero. Therefore, to determine the spin connection in terms of the vielbeins, we only solve
\begin{eqnarray}
    &&T^A_{\ \m\n} \mathcal{E}^{ \m}_{\ C}\mathcal{E}^{ \n}_{\ B}=0,\quad T^A_{\ \m\n} \mathcal{E}^{ \m}_{\ C}\mathcal{E}^{ \n}_{\ b}=0,\quad T^A_{\ \m\n} \mathcal{E}^{ \m}_{\ c}\mathcal{E}^{ \n}_{\ b}=0,\notag\\
    &&T^a_{\ \m\n}\mathcal{E}^{ \m}_{\ c}\mathcal{E}^{ \n}_{\ b}=0,\quad T_{a\m\n}\mathcal{E}^\n_{\ b}\mathcal{E}^\m_{\ C}-T_{b\m\n}\mathcal{E}^\n_{\ a}\mathcal{E}^\m_{\ C}=0.
\end{eqnarray}
However, we note that the following projections: $\Omega_{\mu\ C}^{\ A}\mathcal{E}^\mu_{\ b}$ and $\Omega_{\mu\ b}^{\ A}\mathcal{E}^\mu_{\ A}$ are constrained by only a single equation, namely \cref{projection-spin-connection}. Consequently, the system of equations stays underdetermined and does not uniquely fix these quantities. As a result, infinitely many solutions exist. However, for our purposes, we choose a particular solution that reproduces the affine connection presented in the main text as \cref{SC-connection-Lorentz}\footnote{The rationale for this choice is that the given spin connection components, in the case of Carroll, reduce to the same components computed in Appendix B.2 of ref.~\cite{Hansen:2021fxi}.}.

\medskip

Solving these equations yields the following expressions for the components of spin connection in terms of the vielbeins
\begin{subequations}\label{spin-connection-components}
    \begin{align}
        &\Omega_{\m AB}\mathcal{E}^\m_{\ C}=\frac{1}{2}\left[(d\mathcal{E})_{A\m\n}\mathcal{E}^\m_{\ B}\mathcal{E}^\n_{\ C}+(d\mathcal{E})_{B\m\n}\mathcal{E}^\m_{\ C}\mathcal{E}^\n_{\ A}-(d\mathcal{E})_{C\m\n}\mathcal{E}^\m_{\ A}\mathcal{E}^\n_{\ B}\right],\\
        &\Omega_{\m ab}\mathcal{E}^\m_{\ c}=\frac{1}{2}\left[(d\mathcal{E})_{a\m\n}\mathcal{E}^\m_{\ b}\mathcal{E}^\n_{\ c}+(d\mathcal{E})_{b\m\n}\mathcal{E}^\m_{\ c}\mathcal{E}^\n_{\ a}-(d\mathcal{E})_{c\m\n}\mathcal{E}^\m_{\ a}\mathcal{E}^\n_{\ b}\right],\\
        &\Omega_{\m A C}\mathcal{E}^\m_{\ b}=\frac{1}{2}(d\mathcal{E})_{A \m\n}\mathcal{E}^\m_{\ C}\mathcal{E}^\n_{\ b} + \frac{1}{2}(d\mathcal{E})_{C\,\m\n}\mathcal{E}^{\n}_{ \, A}\mathcal{E}^\m_{\ b}   ,\\
        &\Omega_{\m A b}\mathcal{E}^\m_{\ C}=\frac{1}{2}(d\mathcal{E})_{A \m\n}\mathcal{E}^\n_{\ C}\mathcal{E}^\m_{\ b}+ \frac{1}{2}(d\mathcal{E})_{C\,\m\n}\mathcal{E}^{\n}_{\ A}\mathcal{E}^\m_{\ b}   ,\\
        &\Omega_{\m A b}\mathcal{E}^\m_{\ d}=\frac{1}{2}(d\mathcal{E})_{A \m\n}\mathcal{E}^\n_{\ d}\mathcal{E}^\m_{\ b},\\
        &\Omega_{\m ab}\mathcal{E}^\m_{\ C}=\frac{1}{2}\left[(d\mathcal{E})_{b\m\n}\mathcal{E}^\n_{\ a}\mathcal{E}^\m_{\ C}-(d\mathcal{E})_{a\m\n}\mathcal{E}^\n_{\ b}\mathcal{E}^\m_{\ C}\right].
    \end{align}
\end{subequations}
We are now all set to use the vielbein postulate  \cref{vielbein-postulates} to compute the affine connection $\widehat C^\rho_{\ \m\n}$. However we keep in mind that $\widehat C^\rho_{\ \m\n}=\widehat{C}^\sigma_{\ \m\n}(\mathcal{E}_\sigma^{\ A}\mathcal{E}^\rho_{\ A}+\mathcal{E}_\sigma^{\ a}\mathcal{E}^\rho_{\ a})$. We therefore contract postulates \cref{postulate-1} and \eqref{postulate-2} with $\mathcal{E}^\sigma_{\ A}$ and $\mathcal{E}^\sigma_{\ a}$, respectively and then add them. This gives the following result
\begin{equation}
    \widehat C^\sigma_{\ \m\n}=\mathcal{E}^\sigma_{\ A} \del_\m \mathcal{E}_{ \n}^{\ A}+\mathcal{E}^\sigma_{\ a}\del_\m \mathcal{E}_{ \n}^{\ a}+\Omega_{\m \ B }^{\ A}\mathcal{E}^{\ B}_{ \n}\mathcal{E}^\sigma_{\ A} + \Omega_{\m \ b }^{\ A}\mathcal{E}^{\ b}_{ \n}\mathcal{E}^\sigma_{\ A}+\Omega_{\m \ b }^{\ a}\mathcal{E}^{\ b}_{ \n}\mathcal{E}^\sigma_{\ a} \,.
\end{equation}
Now, using \cref{spin-connection-components} it can be shown
\begin{align}
    \notag \widehat C^\rho_{\ \mu\nu}=&\frac{V^{\rho\lambda}}{2}\left( \partial_\mu V_{\lambda\nu} + \partial_\nu V_{\mu\lambda} - \partial_\lambda V_{\mu\nu} \right)+\frac{\Pi^{\rho\lambda}}{2}\left( \partial_\mu \Pi_{\lambda\nu} + \partial_\nu \Pi_{\mu\lambda} - \partial_\lambda \Pi_{\mu\nu} \right)\\&\quad +\frac{\Pi^{\rho\lambda} V_{\nu\sigma}V^{\sigma\alpha}}{2}\left( \partial_\mu \Pi_{\alpha\lambda} + \partial_\lambda \Pi_{\mu\alpha} - \partial_\alpha \Pi_{\mu\lambda} \right).
\end{align}
We have thus reproduced the connection \cref{SC-connection-Lorentz} presented in the main text. This connection in the limit $\epsilon\to 0$ gives
\begin{eqnarray}
     \notag \widetilde C^\rho_{\ \mu\nu}=&&\frac{v^{\rho\lambda}}{2}\left( \partial_\mu v_{\lambda\nu} + \partial_\nu v_{\mu\lambda} - \partial_\lambda v_{\mu\nu} \right)+\frac{h^{\rho\lambda}}{2}\left( \partial_\mu h_{\lambda\nu} + \partial_\nu h_{\mu\lambda} - \partial_\lambda h_{\mu\nu} \right)\\&&\quad +\frac{h^{\rho\lambda}v_{\nu\sigma}v^{\sigma\alpha}}{2}\left( \partial_\mu h_{\alpha\lambda} + \partial_\lambda h_{\mu\alpha} - \partial_\alpha h_{\mu\lambda} \right).
\end{eqnarray}
This connection is compatible with a generic unified geometry because we did not set all the torsion components to zero, which would otherwise impose constraints on the geometry.

\medskip 

We must also highlight that the unified compatible connection is not invariant under local unified transformations. Instead, it transforms as
\begin{eqnarray}
    \delta_\lambda \widetilde C^\rho_{\ \m\n}=v^{\rho\lambda}\varpi_{\lambda \m\n}+\varpi^{\rho\alpha}(h_{\alpha\m\n}+v_{\n\sigma}v^{\sigma\lambda}h_{\lambda \m\alpha})+h^{\rho\alpha}\varpi_{\n\sigma}v^{\sigma\lambda}h_{\lambda \m\alpha}\neq 0.
\end{eqnarray}
where we have used the notation in \cref{notation-1} and
\begin{subequations}
    \begin{align}
        &\varpi_{\m\n}=\lambda^A_{\ c}\eta_{AB}(e^{\ B}_\m e^{\ c}_\n+e^{\ B}_\n e^{\ c}_\m),\quad  \varpi^{\m\n}=-\lambda^B_{\ a}\eta^{ab}(e^\m_{\ B}e^\n_{\ b}+e^\n_{\ B}e^\m_{\ b}),\\
        &\varpi_{\lambda\m\n}=\frac{1}{2}\left(\del_\m \varpi_{\lambda \n}+\del_\n \varpi_{\m\lambda}-\del_\lambda \varpi_{\m \n}\right).
    \end{align}
\end{subequations}
Therefore, this connection depends on the choice of a local observer. However, the compatibility conditions \cref{SC-compatibility} remain satisfied because it can be shown: $v^{\n\sigma}\delta_\lambda \widetilde C^\rho_{\ \m\n}+v^{\n\rho}\delta_\lambda \widetilde C^\sigma_{\ \m\n}=0$ and $h_{\rho\sigma}\delta_\lambda \widetilde C^\rho_{\ \m\n}+h_{\rho\nu}\delta_\lambda \widetilde C^\rho_{\ \m\sigma}=0$.
\section{Mapping unified connection}
\label{app:mapping connection}
In this appendix, we explicitly show how we derive various versions \cref{PNR-connection,PUL-connection} of the affine connection \cref{SC-connection-Lorentz}, by fixing $s$ and $n$. We then take the $\epsilon\to0$ limit to obtain the connections compatible with the Galilean and Carroll geometries.

\medskip

\underline{\textit{Galilean}:} We fix $(s,n)=(0,d-1)$, for which the redefined variables ($V^{\m\n},V_{\m\n},\Pi_{\m\n}$ and $\Pi^{\mu\nu}$) reduce to the forms given in \cref{galilean-vielbeins}. Substituting these into \cref{SC-connection-Lorentz}, we obtain
\begin{align}
    \widehat C^\rho_{\m\n}=&V^{\rho\s}V_{\s\m\n}+\underset{\circled{1}}{\underbrace{\frac{\Pi^\rho\Pi^{\s}}{2}\left(\del_{\m}(\Pi_{\s}\Pi_{\n})+\del_{\n}(\Pi_{\s}\Pi_{\m})-\del_{\s}(\Pi_{\m}\Pi_{\n})\right)}}\\
    &+\underset{\circled{2}}{\underbrace{\frac{\Pi^\rho\Pi^{\alpha}}{2} V_{\n \sigma}V^{\sigma \lambda}\left(\del_{\m}(\Pi_{\l}\Pi_{\alpha})+\del_{\alpha}(\Pi_{\l}\Pi_{\m})-\del_{\l}(\Pi_{\m}\Pi_{\alpha})\right)}}.\notag
\end{align}
Now, let's simplify the individual terms $\circled{1}$ and $\circled{2}$. First term reduces as
\begin{align}
    \circled{1}&=\frac{\Pi^\rho\Pi^{\s}}{2}\left(\del_{\m}(\Pi_{\s}\Pi_{\n})+\del_{\n}(\Pi_{\s}\Pi_{\m})-\del_{\s}(\Pi_{\m}\Pi_{\n})\right),\\
    &=-\Pi^{\rho}\del_{(\m} \Pi_{\n)}+\Pi^\rho\Pi^{\s}\Pi_{(\m}\del_{\n)}\Pi_\s-\frac{\Pi^\rho\Pi^{\s}}{2}\del_\s(\Pi_{\m}\Pi_{\n}).\notag
\end{align}
Next term can be simplified as
\begin{align}
    \circled{2}&=\frac{\Pi^\rho\Pi^{\alpha}}{2} V_{\n \sigma}V^{\sigma \lambda}\left(\del_{\m}(\Pi_{\l}\Pi_{\alpha})+\del_{\alpha}(\Pi_{\l}\Pi_{\m})-\del_{\l}(\Pi_{\m}\Pi_{\alpha})\right),\\
    &=\notag \frac{\Pi^\rho\Pi^{\alpha}}{2} V_{\n \sigma}V^{\sigma \lambda}\left(\Pi_{\alpha}\del_{\m}(\Pi_{\l})+\Pi_{\m}\del_{\alpha}(\Pi_{\l})-\del_{\l}(\Pi_{\m}\Pi_{\alpha})\right),\\
    &=\notag \frac{\Pi^\rho\Pi^{\alpha}}{2}(\delta^\lambda_{\ \n}+\Pi^\lambda \Pi_{\n})\left(\Pi_{\alpha}\del_{\m}(\Pi_{\l})+\Pi_{\m}\del_{\alpha}(\Pi_{\l})-\del_{\l}(\Pi_{\m}\Pi_{\alpha})\right),\\
    &=\notag \frac{\Pi^\rho}{2}\Pi^\alpha  \del_\alpha (\Pi_{\n}\Pi_{\m})-\Pi^\rho\Pi^\alpha \Pi_{(\m} \del_{\n)} \Pi_\alpha -\Pi^\rho \del_{[\m}\Pi_{\n]}.
\end{align}
Now, by adding these two contributions, we obtain
\begin{eqnarray}
    \circled{1}+\circled{2}=-\Pi^\rho \del_\m \Pi_{\n}.
\end{eqnarray}
Therefore, the affine connection \cref{SC-compatibility-Lorentzian} reduces to 
\begin{eqnarray}
    \widehat C^\rho_{\ \m\n}=-\Pi^\rho \del_{\m}\Pi_\n+V^{\rho\s}V_{\s \m\n}.
\end{eqnarray}
In the limit $\epsilon\to0$, we get a connection compatible with the Galilean geometry
\begin{eqnarray}
    \lim_{\epsilon\to0}\widehat C^\rho_{\ \m\n}=\widetilde C^\rho_{\ \m\n}=-h^\rho\del_\m h_\n +v^{\rho\s}v_{\s \m\n}.
\end{eqnarray}
\underline{\textit{Carroll}:} Now we fix $(s,n)=(1,1)$, and consequently, the redefined variables can be written as \cref{redefined-c}. On substituting it in \cref{SC-compatibility-Lorentzian}, we get
\begin{align}
    \widehat{C}^{\rho}_{\ \m\n}=\underset{\circled{1}}{\underbrace{\frac{V^\rho V^{\lambda}}{2}(\del_\m (V_\l V_\n)+\del_\n (V_\l V_\m)-\del_\l (V_\m V_\n))}}+\Pi^{\rho\alpha }\Pi_{\alpha \m\n}-\underset{\circled{2}}{\underbrace{\Pi^{\rho\alpha }V_\n V^{\l}\Pi_{\lambda \m\a}}}.
\end{align}
It is easy to see that
\begin{eqnarray}
    V^\l \Pi_{\l \m\a}=\frac{V^\l}{2}(\del_{\m}\Pi_{\l \a}+\del_{\a}\Pi_{\l \m}-\del_{\l}\Pi_{\m \a})=-\frac{1}{2}\pounds_V \Pi_{\m\a}=K_{\m\a }.
\end{eqnarray}
Therefore, the term $\circled{2}$=$\Pi^{\rho \a}V_\n K_{\m\a}$. We now simplify the first term
\begin{align}
    \circled{1}&=\frac{V^\rho V^{\lambda}}{2}(\del_\m (V_\l V_\n)+\del_\n (V_\l V_\m)-\del_\l (V_\m V_\n)),\\
    &=\notag - V^\rho \del_{(\m} V_{\n)}+V^\rho V^\l  V_{(\m}\del_{\n)}V_\l-\frac{V^\rho V^\l}{2}\del_\l (V_{\m}V_\n),\\
    &=\notag - V^\rho \del_{(\m} V_{\n)}-V^\rho V_{(\m}\pounds_V V_{\n)}
\end{align}
Therefore, the affine connection \cref{SC-compatibility-Lorentzian} becomes
\begin{eqnarray}
    \widehat{C}^\rho_{\ \m\n}=- V^\rho \del_{(\m} V_{\n)}-V^\rho V_{(\m}\pounds_V V_{\n)}+\Pi^{\rho\alpha }\Pi_{\alpha \m\n}-\Pi^{\rho \a}V_\n K_{\m\a}.
\end{eqnarray}
Now, in the limit $\epsilon\to0$, we get a Carroll compatible connection 
\begin{eqnarray}
    \lim_{\epsilon\to0}\widehat C^\rho_{\ \m\n}=\widetilde C^{\rho}_{\ \m\n}=- v^\rho \del_{(\m} v_{\n)}-v^\rho v_{(\m}\pounds_v v_{\n)}+h^{\rho\alpha }h_{\alpha \m\n}-h^{\rho \a}v_\n \mc{K}_{\m\a},
\end{eqnarray}
where we have used the definition $\mc{K}_{\m\n}=-\frac{1}{2}\pounds_v h_{\m\n}$, same as defined in the main text.
\section{Fiber bundle structure of unified spacetime}
\label{app:fibre bundle}
In general, the unified geometry described by \cref{leading-order-variables} endows the spacetime manifold with a fiber bundle structure. Therefore, locally, spacetime has the following structure
\begin{eqnarray}
    \mathcal{M}\simeq\underset{\text{base}}{\mathcal{B}}\times\underset{\text{fiber}}{\mathcal{F}}.
\end{eqnarray}
In our construction, the dimensionality of the base space and the fiber is controlled by the parameter `$n$'. The Lorentzian and Euclidean classes are extreme cases of this feature. In these cases, the dimensionality of either the fiber or the base space becomes zero. Hence, we shift our focus to Galilean and Carrollian classes (equivalently, $p$-Galilei and $p$-Carroll branes).\\
 \newline
 $\circledast$\textit{For this appendix, we use $\{T,Z\}$ and $\{i,j\}$ indices for the fiber and base coordinates, respectively.}\\
 \newline
 Let's consider a $d$-dimensional manifold with unified geometry that contains an $ n$-dimensional fiber and a $(d-n) $ dimensional base. As we know, the bundle structure contains a projection map $\pi:\mathcal{M}\to\mathcal{B}$, it further induces a linear surjective map between the corresponding tangent bundle, given by $(\text{d}\pi)_{\mathbf{x}}:T_{\mathbf{x}}\mathcal{M}\to T_{\mathbf{x}}\mathcal{B}$. 

\medskip

\begin{figure}[h!]
    \centering
    \includegraphics[width=0.8\linewidth]{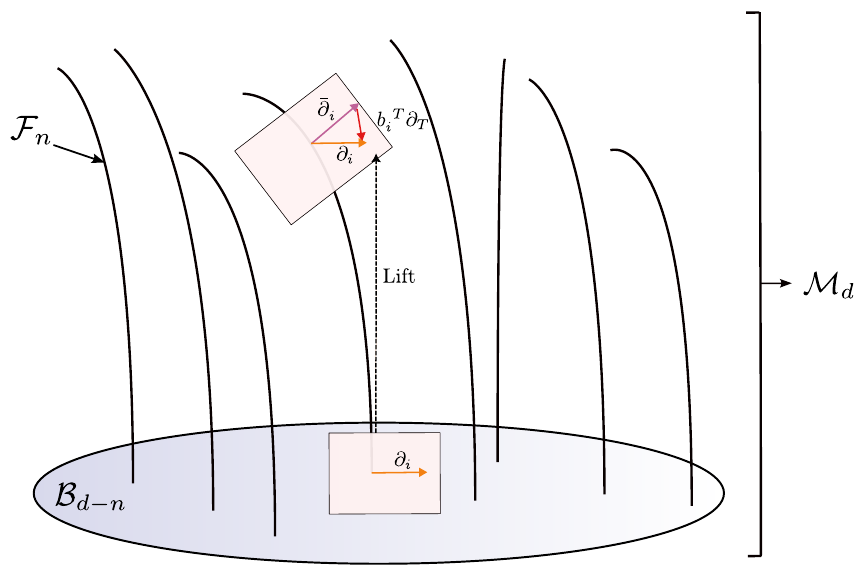}
    \caption{Pictorial illustration of the deviation in the vertical lift of the horizontal vector $\del_i$.}
    \label{fig:placehold1}
\end{figure}

The vertical subspace at a point $\mathbf{x}=(x^T, x^i)$ is defined by the bundle 
structure as $\mathcal{V}_\mathbf{x}\mathcal{M} = \ker\left(\text{d}_\mathbf{x}\pi \right)$, i.e. the set 
of all tangent vectors that project to zero on $\mathcal{B}$. However, the tangent 
space admits a decomposition as
\begin{eqnarray}
    T_\mathbf{x}\mathcal{M} \cong \mathcal{V}_\mathbf{x}(\mathcal{M}) 
    \oplus \mathcal{H}_\mathbf{x}(\mathcal{M}),
\end{eqnarray}
where the horizontal subspace $\mathcal{H}_\mathbf{x}(\mathcal{M})$ is \textit{not} 
canonically fixed by the bundle structure alone. It requires additional data to fix a horizontal section. An \textit{Ehresmann connection} is precisely this choice. It assigns to each point 
$\mathbf{x} \in \mathcal{M}$ a horizontal subspace $\mathcal{H}_\mathbf{x} (\mathcal{M})$ 
that varies smoothly across the manifold and induces the following isomorphism
\begin{eqnarray}
    \text{d}\pi\big|_{\mathcal{H}_\mathbf{x}\mathcal{M}} : \mathcal{H}_\mathbf{x}(\mathcal{M} )
    \xrightarrow{\quad \cong \quad} T_{\pi(\mathbf{x})}\mathcal{B} \, .
\end{eqnarray}
%
In this chart, the connection is encoded in the components 
$b_i^{\ T}(\mathbf{x})$, which specify how a base vector  
$\mathrm{V} = \mathrm{V}^i\del_i \in T_{\pi(\mathbf{x})}\mathcal{B}$ gets a lift to a horizontal 
vector  in $T_\mathbf{x}\mathcal{M}$,
\begin{eqnarray}
    \bar{\mathrm{V}} = \mathrm{V}^i\del_i - \mathrm{V}^ib_i^{\ T}(\mathbf{x})\del_T.
\end{eqnarray}
The second term precisely encodes the failure of $\del_i$ to be horizontal on its own, 
and $b_i^{\ T}$ measures this deviation (see \cref{fig:placehold1}). 

\medskip

Now we have all the ingredients to describe a manifold with unified geometry. In this coordinate chart, the unified geometry has a block diagonal decomposition 
\begin{eqnarray}\label{app2-1}
    h_{\m\n}=\begin{pmatrix}
        0&0\\
        0& h_{ij}
    \end{pmatrix},\quad  v^{\m\n}=\begin{pmatrix}
        v^{TZ}&0\\
        0&0
    \end{pmatrix}.
\end{eqnarray}
Consequently, the inverse data is given by
\begin{subequations}\label{app2-2}
    \begin{align}
         &v_{\m\n}dx^\m dx^\n=v_{TZ}dx^Tdx^Z+2b_{iT}dx^Tdx^i+b_{iT}b_{j}^{\ T}dx^idx^j,\\
        &h^{\m\n}\del_\m \del_\n=b^T_{\ i}b^{iZ}\del_T\del_Z-2b^{iT}\del_T \del_i+h^{ij}\del_i\del_j.
    \end{align}
\end{subequations}
where all capital/lowercase Latin indices are raised and lowered with the inverse pairs $(v_{TZ},v^{TZ})$ and $(h^{ij},h_{ij})$\footnote{We have to be careful with this because $h_{\m\n}$ and $v^{\m\n}$ are degenerate, hence they do not have an inverse. However, their non-degenerate blocks, $h_{ab}$ and $v^{AB}$, can be inverted as a lower-dimensional matrix.}, respectively. It can be checked that \cref{app2-1} with \cref{app2-2} satisfies all conditions in \cref{LO-ortho-norm} and hence qualifies to be a consistent data set for unified geometry. Further, using \cref{local-unified-transformation} we can determine the variation of the Ehresmann connection under local unified transformation
\begin{eqnarray}
    \delta_\lambda b^{iT}=2\eta^{cd}\lambda^B_{\ c}e^{(i}_{\ B}e^{T)}_{\ d}.
\end{eqnarray}
Therefore, we could always go to a local frame where $b^T_{\ i}=0$. In this frame, the unified geometry reduces to
\begin{subequations}\label{local-frame-choice}
    \begin{align}
         &v^{\m\n}\del_\m \del_\n=v^{TZ}\del_T\del_Z,\quad h_{\m\n}dx^\m dx^\n =h_{ij}dx^i dx^j,\\
         &v_{\m\n}dx^\m dx^\n=v_{TZ}dx^T dx^Z,\quad h^{\m\n}\del_\m \del_\n=h^{ij}\del_i\del_j.
    \end{align}
\end{subequations}
Although the choice in \cref{local-frame-choice} breaks the local symmetries, it is often convenient for computing quantities that are invariant under these symmetries, since the chosen data appear only through invariant combinations. The final results remain independent of this choice. Examples of this subtlety are discussed in the main text in the context of the string Carroll geometries, see \cref{near-kerr-1} and \eqref{near-kerr-2}, where a particular choice of inverse data is made to simplify the analysis, exploiting the fact that the gravity equations are invariant under local string Carroll symmetries.

\section{Gauging of unified algebra}
\label{app:Gauging}
In this appendix, we discuss the gauging of the unified algebra \cref{unified-algbera} and derive a dictionary of transformations together with several useful geometric quantities in the spirit of a first principles approach. We further show that this dictionary agrees with the results obtained in the main text via the limiting procedure. In order to gauge the unified algebra we follow the procedure used in \cite{Andringa:2010it,Bergshoeff:2014jla,Hartong:2015zia,Hartong:2015xda}. The Galilean algebra and its central extension, the Bargmann algebra, were discussed in \cite{Andringa:2010it,Hartong:2015zia} and \cite{Bergshoeff:2014jla,Hartong:2015xda} covers the gauging of Carroll algebra.\\
\newline
We start with defining a 1-form connection of the gauge group
\begin{eqnarray}
    \mathcal{A}_\m=e^{\ a}_{ \m}\widetilde P_a+e^{\ A}_{ \m}\widetilde P_A+\Omega^{\ Aa}_{ \m}\widetilde J_{aA}+\frac{1}{2}\Omega^{\ AB}_{ \m}\widetilde J_{AB}+\frac{1}{2}\Omega^{\ ab}_{ \m}\widetilde J_{ab}.
\end{eqnarray}
Here, $e^{\ A}_{ \mu}$ and $e^{\ a}_{ \mu}$ are the gauge fields associated with local translations, which admit an interpretation as vielbeins or equivalently local basis in the geometric viewpoint. On the other hand, $\Omega^{\ Aa}_{ \mu}$, $\Omega^{\ AB}_{ \mu}$, and $\Omega^{\ ab}_{ \mu}$ are the gauge fields corresponding to local boosts and local rotations. However, their precise interpretation is not yet fixed, since the parameters $(s,n)$ remain arbitrary. In the geometric perspective, these fields are known as the spin connections.\\
\newline
This connection transforms infinitesimally in the adjoint as
\begin{eqnarray}\label{adjoint-transform}
    \delta \mathcal{A}_\m=\del_\m \mathcal{U}+[\mathcal{A}_\m,\mathcal{ U}],
\end{eqnarray}
where $\mathcal{U}$ is the local transformation parameter. This, in general, can be written as
\begin{eqnarray}
    \mathcal{U}=\xi^\m \mathcal{A}_\m+\mathcal{J}
\end{eqnarray}
with 
\begin{eqnarray}
    \mathcal{J}=\widetilde J_{aA} \lambda^{Aa}-\frac{1}{2}\widetilde J_{AB} \lambda^{AB}-\frac{1}{2}\widetilde J_{ab} \lambda^{ab}.
\end{eqnarray}
We want to interpret $\xi^\m$ as the parameter for local translations generating diffeomorphisms and $\lambda^{Aa},\lambda^{AC}$ and $\lambda^{ab}$ \footnote{We must be careful with the placement of indices. To match the transformation rules used in the main text, we adopt the conventions
\begin{equation}
\lambda^{Aa}=\eta^{ab}\lambda^{A}{}_{b},\qquad
\lambda^{AB}=\eta^{BC}\lambda^{A}{}_{C},\qquad
\lambda^{ab}=\eta^{bc}\lambda^{a}{}_{c}.
\end{equation}
 } as the remaining local parameters. Therefore, we instead work with the following transformations given below\footnote{Here, we have defined a new local transformation deriving it from the adjoint transformations \cref{adjoint-transform}. However, to be more precise, we are no longer gauging unified algebra from this point onwards, refer to \cite{Figueroa-OFarrill:2022mcy}. But the gauge fields retains its interpretation.} 
\begin{eqnarray}
\bar\delta\mathcal{A}_\m=\delta\mathcal{A}_\m -\xi^\n \mathcal{F}_{\m\n}=\pounds_\xi \mathcal{A}_\m+\del_\m \mathcal{J}+[\mathcal{A}_\m,\mathcal{J}],
\end{eqnarray}
where $\mathcal{F}_{\m\n}$ is the curvature associated with $\mathcal{A}_\m$ defined as
\begin{align}
    \mathcal{F}_{\m\n}&=\del_\m \mathcal{A}_\n -\del_\n \mathcal{A}_\m+[\mathcal{A}_\m,\mathcal{A}_\n]\notag\\
    &=\widetilde P_A R^{\ \ A}_{ \m\n}(\widetilde P_B)+\widetilde P_a R^{\ \ a}_{ \m\n}(\widetilde P_b)+\widetilde J_{aA}R^{\ \ Aa}_{ \m\n}(\widetilde J_{bB})+\frac{1}{2}\widetilde J_{AB}R^{\ \ AB}_{ \m\n}(\widetilde J_{CD})+\frac{1}{2}\widetilde J_{ab}R^{\ \ ab}_{ \m\n}(\widetilde J_{cd}).
\end{align}
Now component wise, $\bar\delta$ transformation acts as
\begin{subequations}
    \begin{align}
        \bar\delta e^{\ a}_{ \m}&= \pounds_\xi e^{\ a}_{ \m} +e_\m^{\ b}\lambda_{\ b}^{ a},\\
        \bar\delta e^{\ A}_{ \m}&=\pounds_\xi e^{\ A}_{ \m}+e_\m^{\ B}\lambda_{\ B}^{ A}+e^{\ a}_\m \lambda_{\ a}^{ A},\\
        \bar\delta \Omega^{\ Aa}_{ \m}&=\pounds_\xi \Omega^{\ Aa}_{ \m}+\del_\m \lambda^{Aa}+\Omega_\m^{\ AB}\lambda^{\ a}_{ B}+\Omega_\m^{\ ab}\lambda_{\ b}^{ A}+\Omega_\m^{\ Ba}\lambda_{\ B}^{ A}+\Omega^{\ A b}_\m \lambda_{\ b}^{ a},\\
        \bar\delta \Omega^{\ AB}_{ \m}&=\pounds_\xi \Omega^{\ AB}_{ \m}-\del_\m \lambda^{AB}+\Omega_\m^{\ AD}\lambda_{\ D}^{ B}-\Omega_\m^{\ BD}\lambda^A_{\ D},\\
        \bar\delta \Omega^{\ ab}_{ \m}&=\pounds_\xi \Omega^{\ ab}_{ \m}-\del_\m \lambda^{ab}+\Omega_\m^{\ ad}\lambda_{\ d}^{ b}-\Omega_\m^{\ bd}\lambda^a_{\ d}.
    \end{align}
\end{subequations}
These transformations exactly maps to the transformations in \cref{local-transformation-unified} derived from the limiting procedure. On a similar note, we compute the expressions for components of the curvatures
\begin{subequations}
    \begin{align}
     R^{\ \ a}_{ \m\n}(\widetilde P_b)&=\del_{[\m}e_{\n]}^{\ a}+\Omega_{\m \ c}^{\ a}e_\n^{\ c}-\Omega^{\  a}_{ \n \ c}e^{\  c}_{ \m},\\
       R^{\ \ A}_{ \m\n}(\widetilde P_B) &=\del_{[\m}e_{\n]}^{\ A}-\Omega_{\m \ c}^{\  A}e_\n^{\ c} +\Omega_{\m \ C}^{\ A}e_\n^{\ C}+\Omega^{\  A}_{ \n \ a}e^{\ a}_{ \m}-\Omega^{\  A}_{ \n \ C}e^{\ C}_{ \m},\\
        R^{\ \ Aa}_{ \m\n}(\widetilde J_{bB})&=\del_{[\m}\Omega_{\n]}^{\ Aa}+\Omega^{\ Ca}_{ \n}\Omega^{\ A}_{ \m\  C}+\Omega^{\  A}_{ \n \ b}\Omega^{\ ab}_{ \m}-\Omega^{\ A}_{ \n \ D}\Omega^{\ Da}_{ \m}-\Omega^{\  a}_{ \n \ b}\Omega^{\ Ab}_{ \m},\\
        R^{\ \ AB}_{ \m\n}(\widetilde J_{CD})&=\del_{[\m}\Omega_{\n]}^{\ AB}+\Omega^{\  A}_{ \n\ C}\Omega^{\ CB}_{ \m}-\Omega^{\  B}_{ \n\ C}\Omega^{\ AC}_{ \m},\\
        R^{\ \ ab}_{ \m\n}(\widetilde J_{cd})&=\del_{[\m}\Omega_{\n]}^{\ ab}+\Omega^{\  a}_{ \n\ c}\Omega^{\ cb}_{ \m}-\Omega^{\  b}_{ \n\ c}\Omega^{\ ac}_{ \m}.
    \end{align}
\end{subequations}
We can further introduce inverse vielbeins, $e_{\ a}^{ \m}$ and $e_{\ A}^{ \m}$, satisfying the following orthogonality and normalization conditions
\begin{eqnarray}
e_{\ A}^{ \m}e^{\ A}_{ \m}=n,\quad e_{\ a}^{ \m}e^{ \ a}_{ \m}=d-n,\quad  e_{ \ A}^{ \m}e^{ \ a}_{ \m}=0   ,\quad e_{ \ a}^{ \m}e^{ \ A}_{ \m}=0,\quad e_{ \ A}^{ \m}e^{\ A}_{ \n}+e_{\ a}^{ \m}e^{\ a}_{ \n}=\delta^\m_{\ \n}.\notag\\
\end{eqnarray}
In the first condition, the factor of `$n$' appears because we are gauging unified algebra and $\eta_{AB}$ is an $n$-dimensional matrix. However, transformations do not depend on this detail. The corresponding transformations can then be determined by demanding that these conditions remain preserved. Consequently, we obtain
\begin{subequations}
    \begin{align}
         &\bar \delta e_{\ A}^{ \m}=\pounds_\xi e_{\ A}^{ \m}-\lambda^B_{\ A}e^\m_{\ B},\\
         &\bar\delta e_{\ a}^{ \m}=\pounds_\xi  e_{\ a}^{ \m}-\lambda^{ B}_{\ a}e_{\ B}^\m-\lambda^b_{\ a}e^\m_{\ b}.
    \end{align}
\end{subequations}
These transformations also agree with the results from the limiting procedure. 
\section{Reducing to Galilean gravity}
\label{appen:galilean gravity}

In this appendix, we show that the leading order equations of motion of Galilean gravity, computed in Eq. (6.24) of \cite{c}, can be reproduced from \cref{SC-gravity-eom 0}, presented in the main text. To match these equations we consider the case $\overset{\{-4\}}{\Lambda} = 0$ throughout this appendix. Moreover, we use \cref{eq: galilean data} and
\begin{gather}
    h_{\mu\nu} = -h_\m h_\n \,,\quad h^{\m\n} = -h^\m h^\n \,, \\
    \mathcal{L}^{(0)} := h^{\rho \alpha} v^{\beta\kappa}v^{\m\n}h_{\kappa\rho \nu}h_{\m\alpha \beta} = \frac 1 4 v^{\rho\lambda} v^{\sigma\alpha}\left( dh \right)_{\rho\sigma} \left( dh \right)_{\lambda\alpha} \,,
\end{gather}
for our required derivation. Moreover, note that the Galilean equations of motion, stated in \cite{c}, are obtained by varying w.r.t $\delta h_\m$ and $\delta v_{\m\n}$, whereas our equations of motion \cref{SC-gravity-eom 0} are computed by varying w.r.t to their inverse pair. Hence, we require the use of the following variational calculi to check the consistency of our argument
\begin{align}
    \mathcal{E}^{(h)}_{\mu\nu}\delta h^{\m\n} &= -2 \mathcal{E}^{(h)}_{\mu\nu} h^\n \delta h^\m \,, \\
    \delta v^{\m\n} &= 2 h^{(\m} v^{\n)\rho} \delta h_\rho - v^{\m\rho} v^{\n\sigma} \delta v_{\rho\sigma} \,, \\
    \delta h^\mu &=  h^\m h^\rho \delta h_\rho - v^{\mu\rho} h^\sigma \delta v_{\rho\sigma} \,.
\end{align}

Starting from \cref{SC-gravity-variation}, we obtain
\begin{align}
    \delta \widetilde S &\approx \frac{1}{32\pi G}\int_\mc M \left[\mc E^{(v)}_{\mu\nu}\delta v^{\mu\nu}+\mathcal{E}^{(h)}_{\mu\nu}\delta h^{\m\n}\right] e~d^{d}x, \nonumber \\
    &= \frac{1}{32\pi G}\int_\mc M \left[ \left( 2 h^{(\mu}v^{\nu)\rho} \delta h_\rho - v^{\mu\rho} v^{\nu\sigma} \delta v_{\rho\sigma}  \right) \mathcal E^{(v)}_{\mu\nu} \right. \nonumber \\
    &\qquad \left.-2\left( h^\mu h^\rho \delta h_\rho - v^{\mu\rho} h^\sigma \delta v_{\rho\sigma} \right) h^\nu \mathcal E^{(h)}_{\mu\nu} \right]e \, d^d x \,, \nonumber \\
    &= -\frac{1}{8\pi G}\int_\mc M \left[ \Xi_{(h)}^{ \ \mu} \delta h_\mu + \frac 1 2 \Xi_{(v)}^{ \ \mu\nu} \delta v_{\mu\nu} \right] e \, d^d x \,,
\end{align}
where the equations of motion read as $\Xi_{(h)}^{ \ \mu} = 0$ and $\Xi_{(v)}^{ \ \mu\nu} = 0$. The expressions for $\Xi_{(h)}$ and $\Xi_{(v)}$ are given below
\begin{subequations}
\begin{align}
    \Xi_{(v)}^{ \ \mu\nu} & = -\frac 1 8 v^{\mu\nu} \left( v^{\alpha\beta} v^{\rho\sigma} \left( dh \right)_{\alpha\rho} \left( dh \right)_{\beta\sigma} \right) + \frac 1 2 v^{\mu\alpha} v^{\nu\beta} v^{\rho\sigma}\left( dh \right)_{\alpha\rho} \left( dh \right)_{\beta\sigma} \,, \\
    \Xi_{(h)}^{ \ \mu} &= \frac 1 8 h^\mu \left( v^{\alpha\beta} v^{\rho\sigma} \left( dh \right)_{\alpha\rho} \left( dh \right)_{\beta\sigma} \right) + \frac 12  h^\alpha \left( dh \right)_{\alpha\beta} v^{\beta\rho} v^{\mu\sigma} \left( dh \right)_{\rho\sigma} \nonumber \\
    &\quad\; + \frac 1 {2 e} \partial_\alpha \left( e \, v^{\alpha\rho} v^{\mu\nu} \left( dh \right)_{\rho\nu} \right) \,.
\end{align}
\end{subequations}

These are exactly the equations of motion derived for leading-order Galilean gravity theory in \cite{c}. Thus, we establish the fact that the equations of motion derived in \cref{subsec: String Carroll gravity} can be reduced to the leading order Galilean gravity equations of motion.

\bibliographystyle{JHEP}
\bibliography{refs}
\end{document}